\documentclass[aps,twocolumn,superscriptaddress,preprintnumbers,showpacs]{revtex4}
\usepackage{amsmath,amssymb,bm,epsfig}

\newcommand\eps{\epsilon}
\renewcommand\d{\partial}
\newcommand\grad{\bm{\nabla}}
\newcommand\p{{\bm{p}}}
\newcommand\q{{\bm{q}}}
\renewcommand\k{{\bm{k}}}
\renewcommand\l{{\bm{l}}}
\newcommand\ep{\varepsilon_\p}
\newcommand\eq{\varepsilon_\q}
\newcommand\ek{\varepsilon_\k}
\newcommand\el{\varepsilon_\l}
\newcommand\Ep{E_\p}
\newcommand\Eq{E_\q}

\newcommand\+{\dagger}
\newcommand\<{\langle}
\renewcommand\>{\rangle}
\newcommand\eF{\varepsilon_{\mathrm{F}}}
\newcommand\kF{k_{\mathrm{F}}}
\newcommand\Tr{\mathop{\mathrm{Tr}}}
\newcommand\eb{\varepsilon_\mathrm{b}}
\newcommand\Veff{V_\mathrm{eff}}
\newcommand\muF{{\mu_\mathrm{F}}}
\newcommand\muB{{\mu_\mathrm{B}}}
\newcommand\vs{\bm v_\mathrm s}
\newcommand\ev{\varepsilon_{m\vs}}
\newcommand\Hc{H_\mathrm c}

\begin{document}
\preprint{INT-PUB 06-10, TKYNT-06-10}

\title{Fermi gas near unitarity around four and two spatial dimensions}
\author{Yusuke~Nishida}
\affiliation{Department of Physics, University of Tokyo,
             Tokyo 113-0033, Japan}
\affiliation{Institute for Nuclear Theory, University of Washington,
             Seattle, Washington 98195-1550, USA}
\author{Dam~Thanh~Son}
\affiliation{Institute for Nuclear Theory, University of Washington,
             Seattle, Washington 98195-1550, USA}

\begin{abstract}
 We construct systematic expansions around four and two spatial
 dimensions for a Fermi gas near the unitarity limit.  Near four spatial
 dimensions such a Fermi gas can be understood as a weakly interacting
 system of fermionic and bosonic degrees of freedom.  To the leading and
 next-to-leading orders in the expansion over $\eps=4-d$, with $d$ being
 the dimensionality of space, we calculate the thermodynamic quantities
 and the fermion quasiparticle spectrum, both at unitarity and around
 the unitarity point.  Then the phase structure of the polarized Fermi
 gas in the unitary regime is studied in the $\eps$ expansion.  At
 unitarity the unpolarized superfluid state and a fully polarized normal
 state are separated by a first order phase transition.  However, on
 the BEC side of the unitarity point, in a certain range of the two-body
 binding energy, these two phases are separated in the phase diagram
 by more exotic phases, including gapless superfluid phases with one or
 two Fermi surfaces and a superfluid phase with a spatially varying
 condensate.  We also show that the unitary Fermi gas near two spatial
 dimensions is weakly interacting, and calculate the thermodynamic
 quantities and the fermion quasiparticle spectrum in the expansion over
 $\bar\eps=d-2$.
\end{abstract}

\date{July 2006}

\maketitle

\section{Introduction}
Two-component Fermi gas with zero-range interaction at infinite
scattering length~\cite{Leggett,Nozieres}, frequently referred to as the
unitary Fermi gas, has attracted intense attention across many subfields
of physics.  Experimentally, the system can be realized in atomic traps
using the Feshbach resonance and has been extensively 
studied~\cite{OHara,Jin,Grimm,Ketterle,Thomas,Salomon,Thomas05}.
Since the fermion density is the only dimensionful scale of the unitary
Fermi gas, its properties are universal, i.e., independent of details of
the interparticle interaction.  The unitary Fermi gas is an idealization
of dilute nuclear matter and may be relevant to the physics of neutron
stars~\cite{Bertsch}.  It has been also suggested that its understanding
may be important for high-$T_\mathrm{c}$ superconductivity~\cite{highTc}.

The austere simplicity of the unitary Fermi gas implies great
difficulties for theoretical treatment, because there seems to be no 
parameter for a perturbation theory.  The usual Green's function
techniques for the many-body problem are completely unreliable here
since the expansion parameter $a\kF$ becomes infinite in the unitarity
limit.  Considerable progress has been made by Monte Carlo 
simulations~\cite{Carlson2003,Chen:2003vy,Astrakharchik2004,Carlson:2005kg},
however these simulations have many limitations, which become especially
evident for systems with a population imbalance between two fermion
components (finite polarization). 

Recently, we have proposed a new approach for the unitary Fermi gas
based on the systematic expansion in terms of the dimensionality
of space~\cite{nishida06}, utilizing the specialty of \textit{four} 
spatial dimensions in the unitarity limit~\cite{nussinov04}.  
In this approach, one would extend the problem to arbitrary spatial
dimension $d$ and consider $d$ is close to, but below, four.  Then one
performs all calculations treating $\eps=4-d$ as a small parameter of
the perturbative expansion.  Results for the physical case of three
spatial dimensions are obtained by extrapolating the series expansions
to $\eps=1$.

We used this $\eps$ expansion around four spatial dimensions to
calculate thermodynamic quantities and fermion quasiparticle spectrum in
the unitarity limit and found results which are quite
consistent with those obtained by Monte Carlo simulations and
experiments~\cite{nishida06}.  Very recently, this expansion has been 
successfully applied to atom-dimer and dimer-dimer scatterings in
vacuum~\cite{Rupak:2006jj}.  Thus there are compelling reasons to hope 
that the limit $d\to4$ is not only theoretically interesting but also
practically useful, despite the fact that the expansion parameter $\eps$
is one at $d=3$.  One may hope that the expansion over $4-d$ will be
as fruitful for the Fermi gas at unitarity as it has been in the theory 
of the second order phase transition~\cite{WilsonKogut}. 

This paper has four main purposes.  We will give a detailed account of
the $\eps$ expansion for the unitary Fermi gas studied in our recent
Letter~\cite{nishida06}, extend the results to the Fermi gas with large
but finite scattering length, study a phase diagram of 
a polarized Fermi gas near the unitarity limit, and develop a
complementary expansion for the unitary Fermi gas around two spatial
dimensions. 

The structure of the paper is as follows.  We start with the study of
two-body scattering in vacuum in the unitarity limit for arbitrary
spatial dimension $2<d<4$.  This study will clarify why systematic
expansions around four and two spatial dimensions are possible for the
unitary Fermi gas (Sec.~\ref{sec:vacuum}).  The $\eps$ expansion is
developed for an unpolarized Fermi gas both at unitarity and at large,
but finite, scattering length in Sec.~\ref{sec:4d}.  Then we apply the
$\eps$ expansion to the unitary Fermi gas with unequal densities for the
two fermion components.  Such a system will be called ``polarized''
Fermi gas and has been recently realized in
experiments~\cite{Ketterle-polarized,Hulet-polarized,Ketterle-polarized2,Ketterle-polarized3}.
The phase structure of the polarized Fermi gas in the unitary regime is 
investigated based on the $\eps$ expansion in
Sec.~\ref{sec:polarization}.  We also show in Sec.~\ref{sec:2d} that
there exists a systematic expansion for the Fermi gas in the unitarity
limit around \textit{two} spatial dimensions.  In
Sec.~\ref{sec:matching}, we make an exploratory discussion to connect
the two expansions around four and two spatial dimensions.
The summary and concluding remarks are given in Sec.~\ref{sec:summary}.

\section{Two-body scattering in vacuum \label{sec:vacuum}}

\subsection{Qualitative discussion}
Nussinov and Nussinov~\cite{nussinov04} were the first two to recognize
the special role of two and four spatial dimensions in the unitarity
limit.  Their arguments are very simple and we review them here for
completeness.

In low dimensions $d\leq2$, any attractive potential possesses at
least one bound state.  Therefore, the threshold of the appearance
of the first two-body bound state corresponds to the zero coupling.
It follows that the Fermi gas in the unitary limit corresponds to a
noninteracting Fermi gas at $d\to2$ and the energy per particle
approaches that of the free Fermi gas in this limit.  If one defines a
parameter $\xi$ as the ratio of the energy density of the Fermi gas at
unitarity and that of the free Fermi gas with the same density, 
\begin{equation}
  \xi = \frac E{E_\mathrm{free}},
\end{equation}
then according to the above argument $\xi\to 1$ as $d\to2$ from above.
(The singular character of $d=2$ was also recognized in the earlier
work~\cite{randeria-2d}.)

Now we consider dimensions close to four.  At infinite scattering
length, the wave function of two fermions with opposite spin behaves like
$R(r)=1/r^{d-2}$ at small $r$, where $r$ is the separation between two
fermions.  Therefore, the normalization integral of the wave function
takes the form
\begin{equation}
 \int\!d\bm{r}R(r)^2=\int_0dr\frac{1}{r^{d-3}},
\end{equation}
which has a singularity at $r\to0$ in high dimensions $d\geq4$.  
For these dimensions the two-body wave function has infinite probability
weight at the origin, and the fermion pair looks like a pointlike
boson.  From this observation, Nussinov and Nussinov concluded that the
unitary Fermi gas at $d\to4$ becomes a noninteracting Bose gas.  In
particular, the energy per particle at fixed Fermi energy goes to zero
as $d\to 4$, because all bosons condense into the zero energy state and
their binding energy is also zero.  According to this argument,
$\xi\to0$ as $d\to 4$ from below. 

This qualitative observation by Nussinov and Nussinov, however, leaves
many questions unanswered:
\begin{enumerate}
 \item How do the bosons interact with each other?
 \item How fast does $\xi$ approach 0 as $d\to 4$?
 \item Can one develop a theory based on an expansion over $\eps=4-d$?
\end{enumerate}
In order to develop a framework that can be used to answer these questions,
we need to review the two-body scattering in a general number of spatial
dimension $d$, with special attention to $d$ near four and two, from the
perspective of the diagrammatic approach.  This consideration will
provide a foundation for the construction of systematic expansions for
the many-body problem of the unitary Fermi gas.

\subsection{Around four spatial dimensions}
One can describe the system of spin-$\frac12$ fermions with
short-range interaction by the following Lagrangian density (here and
below $\hbar=1$):
\begin{equation}\label{eq:L_vacuum}
 \mathcal{L} = \sum_{\sigma=\uparrow,\downarrow} \psi_\sigma^\dagger
  \left(i\d_t+\frac{\grad^2}{2m}\right)\psi_\sigma
  +c_0\,\psi^\dagger_\uparrow\psi^\dagger_\downarrow
  \psi_\downarrow\psi_\uparrow.
\end{equation}
Here $m$ is the fermion mass and $c_0$ is the bare attractive coupling
between two fermions.  The last term in Eq.~(\ref{eq:L_vacuum})
corresponds to the interaction potential
\begin{equation}\label{Vdelta}
  V(\bm{r}) = -c_0 \delta(\bm{r}).
\end{equation}
It is well known for $d=3$ that in order to have a finite scattering
length, the potential in Eq.~(\ref{Vdelta}) has to be regularized: One
chooses a particular shape of the potential (for example, the
square-well shape), and then takes the width of the potential to be zero
simultaneously with adjusting the height so that the zero-energy wave
function approaches a finite limit.  We shall show that it is true for
$d$ between two and four considering the two-body scattering.  

The $T$-matrix of the two-body scattering is given by the infinite
summation of bubble diagrams. As a result, its inverse is
\begin{align}\label{eq:c_0}
  T(p_0,\p)^{-1}&=\frac1{c_0}+i\int\!\frac{dk_0\,d\k}{(2\pi)^{d+1}}
  \frac1{\frac{p_0}2-k_0-\varepsilon_{\frac\p2-\k}+i\delta} \notag\\
  &\qquad\qquad\qquad
  \times\frac1{\frac{p_0}2+k_0-\varepsilon_{\frac\p2+\k}+i\delta} \notag\\
  &=\frac1{c_0}-\int_\k \frac1{2\ek-p_0+\frac\ep2-i\delta},
\end{align}
where $\varepsilon_\p=\p^2/2m$ is the kinetic energy of nonrelativistic
particles and we introduced the shorthand notation
\begin{equation}
  \int_\k \equiv \frac{d\k}{(2\pi)^d}.
\end{equation}
The integration over $\k$ in Eq.~(\ref{eq:c_0}) is ultraviolet divergent
as $\int\!d\k/k^2$ if $d>2$ and needs to be regularized. 

The usual way to do this regularization is to assume that there is an
upper limit $\Lambda$ in the momentum integral of Eq.~(\ref{eq:c_0}),
and to adjust the $\Lambda$ dependence of $c_0$ so that the physics does
not depend on $\Lambda$.  If we add and subtract the value of the
integral at $p_0=\p=0$, we have
\begin{equation}
 \begin{split}
  T(p_0,\p)^{-1} &= \frac1{c_0(\Lambda)} 
  - {\int\limits^\Lambda}_{\!\!\!\k}\,\frac1{2\ek} \\
  &\quad - {\int\limits^\Lambda}_{\!\!\!\k}
  \left(\frac1{2\ek-p_0+\frac\ep2-i\delta}-\frac1{2\ek}\right).
 \end{split}
\end{equation}
The second integral is now $\int\!d\k/k^4$ at large $\k$ and converges
for $d<4$.  Thus for $d<4$ one can set the cutoff to infinity in the
second integral.  The first integral is still ultraviolet divergent
but is independent of $p_0$ and $\p$, and it can be absorbed into the
definition of the renormalized coupling $c_0^\mathrm{ren}$ as follows:
\begin{equation}
  \frac1{c_0^\mathrm{ren}} \equiv \frac1{c_0(\Lambda)} 
  - {\int\limits^\Lambda}_{\!\!\!\k}\,\frac1{2\ek}.
\end{equation}
The zero range limit corresponds to sending $\Lambda\to\infty$ but
keeping the renormalized coupling $c_0^\mathrm{ren}$ finite by an
appropriate choice of the bare coupling $c_0(\Lambda)$.  In this limit
all physical quantities should depend only on $c_0^\mathrm{ren}$ but not
on $c_0$ and $\Lambda$ separately.  From the above discussion it is
clear that this limit can be taken only for $d<4$; for $d\geq4$ the
divergences in the fermion loop can no longer be compensated by the
renormalization of one coupling constant $c_0$.  There is no universal
zero range and large scattering length limit for $d\geq4$.

In this paper we will not use the momentum-cutoff regularization
described above.  Rather we shall use the dimensional regularization,
in which the integrals are evaluated in dimensions where they are
ultraviolet finite and then the results are analytically continued to
the dimension $d$.  For example, the integral
\begin{equation}
 \int_\k \frac{\ek^{\,m}}{(\ek^{\,2} + \phi_0^{\,2})^{n/2}}
\end{equation}
converges only for $d+2m<2n$, but we will use the analytical formula
\begin{equation}\label{eq:dimreg}
 \begin{split}
  &\int_\k \frac{\ek^{\,m}}{(\ek^{\,2} + \phi_0^{\,2})^{n/2}}
  \\ & \quad = \frac{\Gamma\left(\frac d4+\frac m2\right)
  \Gamma\left(\frac{n-m}2-\frac d4\right)}
  {2\,\Gamma\left(\frac d2\right) \Gamma\left(\frac n2\right)}
  \left(\frac{m\phi_0}{2\pi}\right)^{d/2} \phi_0^{\,m-n}
 \end{split}
\end{equation}
for all values of $d$.  This regularization scheme is particularly
convenient because the divergent integral in the formula defining the
renormalized coupling $c_0^\mathrm{ren}$ is zero,
\begin{equation} 
 \int_\k\,\frac1{2\ek} = 0 \quad\text{in dimensional\ regularization,}
\end{equation}
and hence the renormalized coupling is just the bare coupling,
$c_0^\mathrm{ren}=c_0$.  In other words, in the dimensional
regularization the subtraction of the leading divergence of the fermion
bubble diagram is implemented automatically.  As long as physical
quantities are insensitive to the ultraviolet cutoff, which is the case 
for $d<4$, the results of the dimensional and momentum-cutoff
regularizations coincide.

\begin{figure}[tp]
 \includegraphics[width=0.48\textwidth,clip]{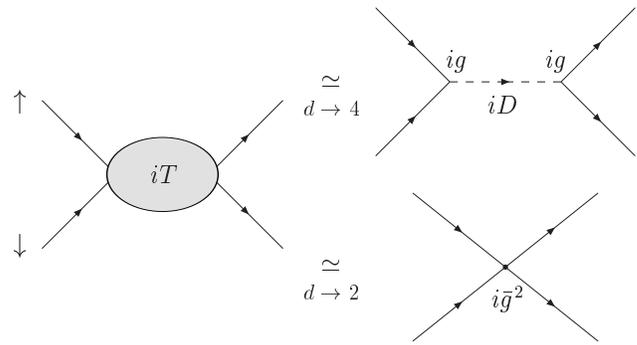}
 \caption{Two-fermion scattering in vacuum in the unitarity limit. The
 $T$-matrix near four spatial dimensions is expressed by the propagation
 of the boson with the small effective coupling $g$, while it reduces to
 a contact interaction with the small effective coupling $\bar g^2$ near
 two spatial dimensions.  \label{fig:scattering}}  
\end{figure}

In the dimensional regularization, the integral in Eq.~(\ref{eq:c_0})
vanishes for $p_0=\p=0$ (scatterings at threshold).  Therefore the limit
of infinite scattering length where $T(0,\bm0)=\infty$ corresponds to
$c_0=\infty$.  In this limit the integration over $\k$ in
Eq.~(\ref{eq:c_0}) can be evaluated explicitly,
\begin{equation}\label{eq:T^-1}
 T(p_0,\p)^{-1}=-\Gamma\!\left(1-\frac d2\right)
  \left(\frac{m}{4\pi}\right)^{\frac d2}
 \left(-p_0+\frac\ep2-i\delta\right)^{\frac d2-1}.
\end{equation}
This expression has a pole at $d=4$ originating from the ultraviolet
divergence of the $\k$ integration.  Thus the $T$-matrix is small when
$d$ is close to four and vanishes at $d=4$, indicating that such a limit
can be thought of as a noninteracting limit.  Substituting $d=4-\eps$
and expanding in terms of $\eps$, the $T$-matrix near four spatial
dimensions becomes
\begin{equation}\label{eq:}
 iT(p_0,\p)=-\frac{8\pi^2\eps}{m^2}\frac{i}{p_0-\frac\ep2+i\delta}+O(\eps^2).
\end{equation}
From this equation it is clear what is special about $d$ close to
four: As the number of dimensions approaches four from below, the
two-fermion scattering amplitude has the form of the scattering process
that occurs through the propagation of an intermediate boson whose mass
is $2m$, as depicted in Fig.~\ref{fig:scattering}.

The propagator of the intermediate boson is
\begin{equation}\label{eq:D_vacuum}
  D(p_0,\p) = \left(p_0 
  - \frac{\ep}2 + i\delta\right)^{-1},
\end{equation}
and the effective coupling of the intermediate boson to the fermion pair
is 
\begin{equation}\label{eq:g}
 g^2=\frac{8\pi^2\eps}{m^2}.
\end{equation}
Then the $T$-matrix, to leading order in $\eps$, can be written as
\begin{equation}\label{eq:T}
 iT(p_0,\p)\simeq(ig)^2 iD(p_0,\p).
\end{equation}
It is natural to interpret the intermediate boson as a bound state of
the two fermions at threshold, which will be referred to simply as a
\textit{boson}.  

An important point is that the effective coupling of the two fermions
into the boson  $g\sim\eps^{1/2}$ is small near four dimensions.  This
fact indicates the possibility to construct a perturbative expansion for
the unitary Fermi gas near four spatial dimensions in terms of the small
parameter $\eps$.

\subsection{Around two spatial dimensions}
Similarly, a perturbative expansion around two spatial dimensions is
possible.  This is because the inverse of the $T$-matrix in
Eq.~(\ref{eq:T^-1}) has another pole at $d=2$ and hence the $T$-matrix
vanishes in this limit, which indicates that $d=2$ from above is again a
noninteracting limit.  Substituting $d=2+\bar\eps$ and expanding in
terms of $\bar\eps$, the $T$-matrix near two spatial dimensions
becomes
\begin{equation}
 iT(p_0,\p)=i\frac{2\pi}m\bar\eps+O(\bar\eps^2).
\end{equation}
If we define the effective coupling at $2+\bar\eps$ dimensions as 
\begin{equation}\label{eq:g-bar}
 \bar g^2=\frac{2\pi}m\bar\eps, 
\end{equation}
then the $T$-matrix to the leading order in $\bar\eps$ can be written as
\begin{equation}
 iT(p_0,\p)\simeq i\bar g^2.
\end{equation}
We see that the $T$-matrix near two spatial dimensions corresponds to
a contact interaction with the small effective coupling 
$\bar g^2\sim\bar\eps$, as depicted in Fig.~\ref{fig:scattering}. 
In this case, the boson propagator $D(p)$ in Eq.~(\ref{eq:T})
is just a constant $-1$.  We defer our discussion of the
expansion over $\bar\eps=d-2$ to Sec.~\ref{sec:2d} and concentrate on
the expansion over $\eps=4-d$.

\subsection{Binding energy of two-body state}
We shall be interested not only in the physics right at the unitarity
point, but also in the vicinity of it.  In other words, $1/c_0$ can be
nonzero in the dimensional regularization.  The case of $c_0<0$
corresponds to the BEC side of the unitarity point, and $c_0>0$
corresponds to the BCS side.  In order to facilitate an extrapolation to
three spatial dimensions, one would like to identify $c_0$ with the
scattering length.  However, this cannot be done directly since the
dimensionality of $c_0$ changes with the number of dimensions. 

One way to circumvent the problem is to parametrize the deviation from
the unitarity by the binding energy of the two-body state.  Clearly,
this method works only on the BEC side of the unitarity point.
Fortunately, as we will find in Sec.~\ref{sec:polarization}, all 
interesting exotic phases of the polarized Fermi gas appear on the
BEC side of the unitarity limit.

The binding energy $\eb$, defined to be positive $\eb>0$, is obtained
from the location of the pole of the $T$-matrix at zero external
momentum: $T(-\eb,\bm0)^{-1}=0$.  From Eqs.~(\ref{eq:c_0}) and
(\ref{eq:T^-1}), we see the coupling $c_0<0$ is related with $\eb$ via
the following equation:
\begin{equation}\label{eq:eb}
 \frac1{c_0}=\Gamma\!\left(1-\frac d2\right)
  \left(\frac{m}{4\pi}\right)^{\frac d2}\eb^{\,\frac d2-1}.
\end{equation}
Near four spatial dimensions, the relationship between $\eb$ and $c_0$
to the leading order in $\eps=4-d$ is given by
\begin{equation}\label{eq:eb-4d}
 \frac1{c_0}\simeq-\frac{\eb}{2\eps}\left(\frac{m}{2\pi}\right)^2.
\end{equation}
In three spatial dimensions, the relationship between $\eb$ and the
scattering length $a=-mc_0/4\pi$ becomes $\eb=(ma^2)^{-1}$.  

At finite density in general $d$ dimensions, we shall use $\eb/\eF$ as
the dimensionless parameter characterizing the deviation from the
unitarity, where $\eF$ is the Fermi energy.  If we recall that the Fermi
energy is related to the Fermi momentum $\kF$ through
$\eF=\kF^{\,2}/(2m)$, one finds the following relationship holds at
$d=3$:
\begin{equation}\label{eq:ebeF}
  \frac\eb\eF = \frac2{(a\kF)^{2}}. 
\end{equation}
We will use these relations to compare results in the expansion over
$\eps=4-d$ with the physics in three spatial dimensions.

\section{Fermi gas near unitarity \label{sec:4d}}

\subsection{Lagrangian and Feynman rules}
Employing the idea of the previous section, we now construct the $\eps$
expansion for the Fermi gas near the unitarity limit. 
We start with the Lagrangian density in Eq.~(\ref{eq:L_vacuum}) and 
introduce two chemical potentials $\mu_\uparrow$ and $\mu_\downarrow$
for the two spin components as follows: 
\begin{equation}
 \mathcal{L} = \sum_{\sigma=\uparrow,\downarrow} \psi_\sigma^\dagger
  \left(i\d_t+\frac{\grad^2}{2m}+\mu_\sigma\right)\psi_\sigma
  +c_0\,\psi^\dagger_\uparrow\psi^\dagger_\downarrow
  \psi_\downarrow\psi_\uparrow.
\end{equation}
After making a Hubbard-Stratonovich transformation, we rewrite the 
Lagrangian density as 
\begin{equation}\label{eq:L}
 \begin{split}
  \mathcal{L} &= \Psi^\+\left(i\d_t + \frac{\sigma_3\grad^2}{2m}
  + \mu\sigma_3 + H\right)\Psi \\ 
  &\qquad\qquad\quad - \frac1{c_0} \phi^*\phi
  + \Psi^\+\sigma_+\Psi\phi + \Psi^\+\sigma_-\Psi\phi^*,
 \end{split}
\end{equation}
where $\Psi=(\psi_\uparrow,\psi^\+_\downarrow)^T$ is a two-component
Nambu-Gor'kov field, and $\sigma_{1,2,3}$ and 
$\sigma_\pm=\frac12(\sigma_1\pm i\sigma_2)$ are the Pauli matrices. 
We define the average chemical potential as
$\mu=(\mu_\uparrow+\mu_\downarrow)/2$ and the chemical potential
difference as $H=(\mu_\uparrow-\mu_\downarrow)/2$. 

The ground state at the finite density system (at least when $H=0$) 
is a superfluid state where
$\phi$ condenses: $\<\phi\>=\phi_0$ with $\phi_0$ being chosen to be 
real.  With that in mind we expand $\phi$ around its vacuum expectation
value $\phi_0$ as 
\begin{equation}\label{eq:coupling}
  \phi=\phi_0+ g\varphi, \qquad g = \frac{(8\pi^2\eps)^{1/2}}m
  \left(\frac{m\phi_0}{2\pi}\right)^{\eps/4}.
\end{equation}
Here we introduced the effective coupling $g\sim\eps^{1/2}$ in
Eq.~(\ref{eq:g}). The extra factor
$\left(m\phi_0/2\pi\right)^{\eps/4}$ was chosen so that $\varphi$
has the correct dimension of a nonrelativistic field~\footnote{The
choice of the extra factor is arbitrary, if it has the correct
dimension, and does not affect the final results because the difference
can be absorbed into the redefinition of the fluctuation field
$\varphi$.  The particular choice of $g$ (or $\bar g$) in
Eq.~(\ref{eq:coupling}) [or Eq.~(\ref{eq:coupling_2d})] will simplify
the form of loop integrals in the intermediate steps.}.  

We now split the Lagrangian into three parts,
$\mathcal{L}=\mathcal{L}_0+\mathcal{L}_1+\mathcal{L}_2$, where
\begin{align}
 \begin{split}\label{eq:L_0}
  \mathcal{L}_0 & = \Psi^\+\left(i\d_t + \frac{\sigma_3\grad^2}{2m}
  + H + \sigma_+\phi_0 + \sigma_-\phi_0\right)\Psi \\ 
  & \qquad + \varphi^*\left(i\d_t+\frac{\grad^2}{4m}\right)\varphi
  - \frac{\phi_0^{\,2}}{c_0}\,,
 \end{split} \\ \notag\\
 \begin{split}\label{eq:L_1}
  \mathcal{L}_1 & = g\Psi^\+\sigma_+\Psi\varphi 
  + g\Psi^\+\sigma_-\Psi\varphi^* + \mu\Psi^\+\sigma_3\Psi \\
  & \qquad + \left(2\mu-\frac{g^2}{c_0}\right)\varphi^*\varphi 
  - \frac{g\phi_0}{c_0}\varphi - \frac{g\phi_0}{c_0}\varphi^*\,, 
 \end{split} \\ \notag\\
 \mathcal{L}_2 & = -\varphi^*\left(i\d_t
 +\frac{\grad^2}{4m}\right)\varphi - 2\mu\varphi^*\varphi\,. \label{eq:L_2}
\end{align}
The Lagrangian density in Eq.~(\ref{eq:L}) does not have the kinetic term
for the boson field $\varphi$.  We add such a term to $\mathcal{L}_0$ and
subtract it in $\mathcal{L}_2$.  Analogously, we add a chemical potential
term for $\varphi$ in $\mathcal{L}_1$ and subtract it in $\mathcal{L}_2$. 
At first sight, the split~(\ref{eq:L_0})--(\ref{eq:L_2}) may appear  
counterintuitive, but it will greatly simplify the organization of the
expansion in powers of $\eps$. 

We shall also see that the condensate $\phi_0$ coincides, to the leading
order in $\eps$, with the energy gap in the fermion quasiparticle
spectrum.  From Eq.~(\ref{eq:eb-4d}), $-g^2/c_0\simeq\eb$ gives the
binding energy of the boson to the leading order in $\eps$ when $c_0$ is
negative.  Throughout this paper, we consider the vicinity of the
unitarity point where $\eb\sim\eps\phi_0$. 

The part $\mathcal{L}_0$ is the Lagrangian density of noninteracting
fermion quasiparticles and bosons with the mass $2m$.  The propagators
of fermion and boson are generated by $\mathcal{L}_0$.  The fermion
propagator $G$ is a $2\times2$ matrix,
\begin{equation}\label{eq:G}
 \begin{split}
  G(p_0,\p) &= \frac1{(p_0+H)^2-E_\p^{\,2}+i\delta} \\
  & \qquad\quad \times
  \begin{pmatrix}
   p_0 + H + \ep & -\phi_0 \\
   -\phi_0 & p_0 + H -\ep
  \end{pmatrix},
 \end{split}
\end{equation}
where $E_\p=\sqrt{\ep^{\,2}+\phi_0^{\,2}}$ is the excitation energy of
the fermion quasiparticle.  
The boson propagator $D$ is given by the same form as in
Eq.~(\ref{eq:D_vacuum}),  
\begin{equation}\label{eq:D}
  D(p_0,\p) = \left(p_0 
  - \frac{\ep}2 + i\delta\right)^{-1}.
\end{equation}

\begin{figure}[tp]
 \includegraphics[width=0.45\textwidth,clip]{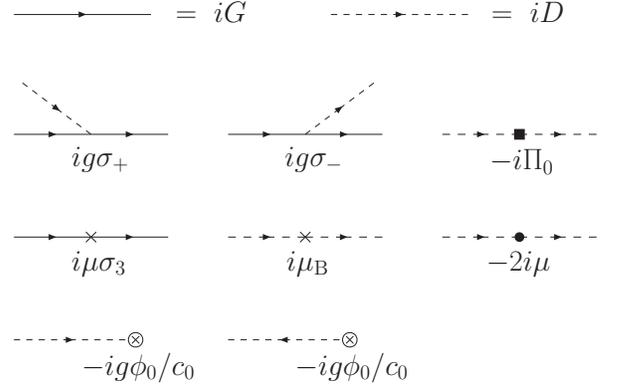}
 \caption{Feynman rules from the Lagrangian density in
 Eqs.~(\ref{eq:L_0})--(\ref{eq:L_2}).  The two vertices in the last
 column come from $\mathcal{L}_2$, while the rest come from
 $\mathcal{L}_1$. Solid (dotted) lines represent the fermion (boson)
 propagator $iG$ $(iD)$.  \label{fig:feynman_rules}}
\end{figure}

The part $\mathcal{L}_1$ generates the vertices depicted in the first
two columns of Fig.~\ref{fig:feynman_rules}.  The first two terms in
$\mathcal{L}_1$ describe the fermion-boson interactions, whose coupling
is proportional to $g\sim\eps^{1/2}$ and small in the limit $\eps\to0$.
This coupling originates from the two-body scattering in vacuum in the
unitarity limit which was studied in the previous section (see
Fig.~\ref{fig:scattering}).  The third and fourth terms are insertions
of chemical potentials, $\mu$ and $\mu_\mathrm{B}\equiv2\mu-g^2/c_0$, to
the fermion and boson propagators, respectively.  We treat these two
insertions as small perturbations for the following reasons.  First, we
shall see that $\mu$ is small compared to the scale set by $\phi_0$,
i.e., $\mu/\phi_0\sim\eps$.  Second, we limit ourselves to a region near
the unitarity limit where the boson binding energy is small compared
to $\phi_0$, i.e., $\eb/\phi_0\sim\eps$ so
$\mu_\mathrm{B}/\phi_0\sim\eps$.  The last two terms in $\mathcal{L}_1$
give tadpoles to the $\varphi$ and $\varphi^*$ fields, which are
proportional to $-ig\phi_0/c_0$.  The condition of cancellation of
tadpole diagrams will determine the value of the condensate $\phi_0$.

Finally, $\mathcal{L}_2$ generates two additional vertices for the boson
propagator as depicted in the last column of
Fig.~\ref{fig:feynman_rules}, which are $-i\Pi_0$ and $-2i\mu$ where
\begin{equation}\label{eq:Pi_0}
  \Pi_0(p_0,\p) = p_0 -\frac{\ep}2.
\end{equation}
These two vertices can be thought of as counterterms to avoid double
counting of certain types of diagrams which are already taken into
$\mathcal{L}_0$ and $\mathcal{L}_1$.  For the rest of this section, 
we will consider the unpolarized case with $H=0$ at zero temperature.

\subsection{Power counting rule of $\epsilon$}
We now develop the power counting rule which allows us to identify
Feynman diagrams that contribute to a given order in $\eps$.  Our
discussion will extend the power counting scheme developed for the 
unitarity limit in Ref.~\cite{nishida06} to the region near the
unitarity where $g^2/|c_0|\sim\eps\phi_0$. 

Let us first consider Feynman diagrams constructed only from
$\mathcal{L}_0$ and $\mathcal{L}_1$, without the vertices from
$\mathcal{L}_2$.  We make a prior assumption $\mu/\phi_0\sim\eps$, which
will be checked later, and consider $\phi_0$ to be $O(1)$.  Each pair of
fermion-boson vertices is proportional to $g^2\sim\eps$ and hence brings
a factor of $\eps$.  Also, each insertion of $\mu\sim\eps$ or
$\muB=2\mu-g^2/c_0\sim\eps$ brings another factor of $\eps$.  Therefore,
the power of $\eps$ for a given diagram is naively $N_g/2+N_\mu$, where
$N_g$ is the number of couplings $g$ from $\mathcal{L}_1$, and
$N_\mu=N_{\muF}+N_{\muB}$ is the sum of the number of $\mu$ insertions
to the fermion line and $\muB$ insertions to the boson line.

However, this naive power counting does not take into account the
possibility of inverse powers of $\eps$ from loop integrals.  Such a
$1/\eps$ behavior arises for integrals which are ultraviolet divergent
at $d=4$.  Let us identify diagrams which have this divergence.

Consider a diagram with $L$ loop integrals, $P_\mathrm{F}$ fermion
propagators, and $P_\mathrm{B}$ boson propagators.  Each momentum loop
integral in the ultraviolet region behaves as
\begin{equation}
 \int dp_0d\p\sim \int d\p\,\ep\sim p^{6},
\end{equation}
while each fermion or boson propagator falls at least as $G(p)\sim
p^{-2}$ or $D(p)\sim p^{-2}$ or faster.  Therefore, the maximal degree
of divergence that the diagram may have is 
\begin{equation}
 \mathcal{D}=6L-2P_\mathrm{F}-2P_\mathrm{B}.
\end{equation}
We call $\mathcal{D}$ the superficial degree of divergence.
$\mathcal{D}=0$ corresponds to the possibility of logarithmic
divergence, $\mathcal{D}=2$ to quadratic divergence, etc. 

Moreover, following an analysis similar to that done in relativistic
field theories~\cite{Peskin:1995ev}, one finds the following relations
which involve the number of external fermion (boson) lines
$E_\mathrm{F(B)}$: 
\begin{equation}\label{eq:loop}
 \begin{split}
  L&=(P_\mathrm{F}-N_\muF)+(P_\mathrm{B}-N_\muB)-N_g+1,\\
  N_g&=(P_\mathrm{F}-N_\muF)+\frac{E_\mathrm{F}}2
  =2(P_\mathrm{B}-N_\muB)+E_\mathrm{B}.
 \end{split}
\end{equation}
With the use of these relations, the superficial degree of divergence
$\mathcal{D}$ can be written as 
\begin{equation}\label{eq:estimate}
 \mathcal{D}=6-2(E_\mathrm{F}+E_\mathrm{B}+N_\mu), 
\end{equation}
which shows that the inverse powers of $\eps$ appear only in
diagrams where the total number of external lines and chemical potential
insertions does not exceed three.  This is similar to the situation in
quantum electrodynamics where infinities occur only in electron and
photon self-energies and the electron-photon triple vertex.

However, this estimation of $\mathcal{D}$ is actually an overestimate:
for many diagrams the real degree of divergence is smaller than given in 
Eq.~(\ref{eq:estimate}).  To see that, we split $G(p)$ into the retarded
and advanced parts: $G(p)=G^\mathrm{R}(p)+G^\mathrm{A}(p)$, where 
$G^\mathrm{R}$ ($G^\mathrm{A}$) has poles only in the lower (upper) half
of the complex $p_0$ plane.  It is easy to see that the ultraviolet
behaviors of different components of the propagators are different:
\begin{align}\label{eq:analytic}
  &G_{11}^\mathrm{R}(p)\sim G_{22}^\mathrm{A}(p)\sim 
  D^\mathrm{R}(p)\sim p^{-2},\phantom{\sum}\\
  &G_{12}(p)\sim G_{21}(p)\sim p^{-4},\quad  \notag
  G_{11}^\mathrm{A}(p)\sim G_{22}^\mathrm{R}(p)\sim p^{-6}.
\end{align}
Note that since the boson propagator $D(p)$ has a pole only on the lower
half plane of $p_0$, we have only the retarded Green's function for the
boson $D^\mathrm{A}(p)=0$. 

\begin{figure}[bp]
 \includegraphics[width=0.45\textwidth,clip]{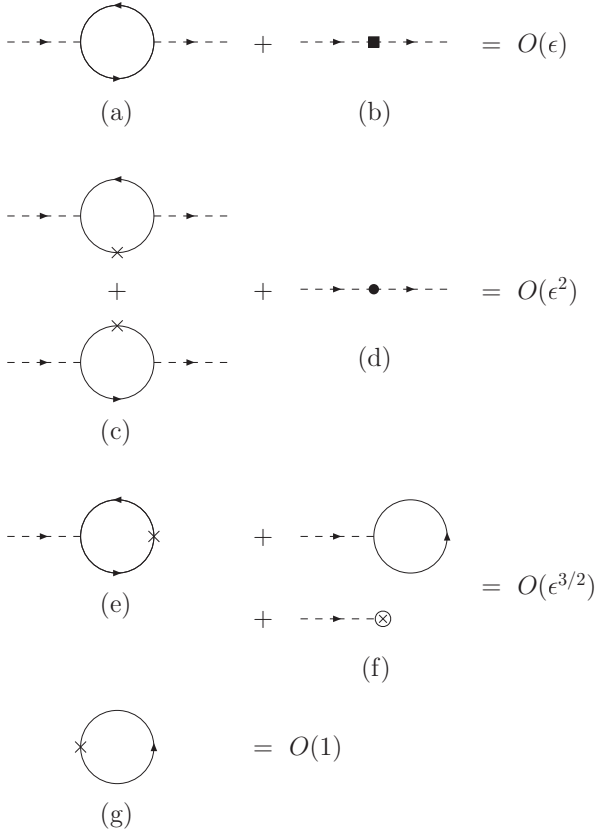}
 \caption{Four apparent exceptions of the naive power counting rule of
 $\eps$ [(a), (c), (e), (g)].  The boson self-energy diagram (a) or (c)
 is combined with the vertex from $\mathcal{L}_2$, (b) or (d), to
 restore the naive $\eps$ power counting.  The tadpole diagram in (e)
 cancels with other tadpole diagrams at the minimum of the effective
 potential. The vacuum diagram (g) is the only exception of the naive
 power counting rule of $\eps$, which is $O(1)$ instead of $O(\eps)$.
 \label{fig:cancel}}  
\end{figure}

From these analytic properties of the propagators in the ultraviolet
region and the vertex structures in $\mathcal{L}_1$ as well as the
relation of Eq.~(\ref{eq:estimate}), one can show that there are only
four skeleton diagrams which have the $1/\eps$ singularity near
$d=4$.  They are one-loop diagrams of the boson self-energy 
[Figs.~\ref{fig:cancel}(a) and \ref{fig:cancel}(c)], the $\varphi$
tadpole [Fig.~\ref{fig:cancel}(e)], and the vacuum diagram
[Fig.~\ref{fig:cancel}(g)].  We shall examine these apparent four
exceptions of the naive power counting rule of $\eps$ one by one. 

The first diagram, Fig.~\ref{fig:cancel}(a), is the one-loop diagram
of the boson self-energy.  The frequency integral can be done
explicitly, yielding
\begin{widetext}
\begin{equation}\label{eq:Pi_a}
 \begin{split}
  -i\Pi_\mathrm{a}(p) &= -g^2\int\!\frac{dk}{(2\pi)^{d+1}}\, 
  G_{11}\!\left(k+\frac p2\right)G_{22}\!\left(k-\frac p2\right) \\ 
  &= ig^2 \int_\k \frac1{4E_{\k-\frac\p2}E_{\k+\frac\p2}}
  \left[\frac{(E_{\k-\frac\p2}+\varepsilon_{\k-\frac\p2})
  (E_{\k+\frac\p2}+\varepsilon_{\k+\frac\p2})}
  {E_{\k-\frac\p2}+E_{\k+\frac\p2}-p_0-i\delta}
  +\frac{(E_{\k-\frac\p2}-\varepsilon_{\k-\frac\p2})
  (E_{\k+\frac\p2}-\varepsilon_{\k+\frac\p2})}
  {E_{\k-\frac\p2}+E_{\k+\frac\p2}+p_0-i\delta}\right].
 \end{split}
\end{equation}
\end{widetext}
The integral over $\k$ is ultraviolet divergent at $d=4$ and has a pole
at $\eps=0$.  Thus $\Pi_\mathrm{a}(p)$ is $O(1)$ by itself instead
of $O(\eps)$ according to the naive power counting.  The residue at
the pole is
\begin{equation}
 \begin{split}
  \Pi_\mathrm{a}(p) & = -g^2\int_\k
  \left( 2\ek - p_0 + \frac{\ep}2 \right)^{-1} + \cdots \\
  &= -\left(p_0-\frac{\ep}2\right) + O(\eps),
 \end{split} 
\end{equation}
and is canceled out exactly by adding the vertex $\Pi_0$ in
$\mathcal{L}_2$.  Therefore the diagram of the type in
Fig.~\ref{fig:cancel}(a), when combined with the vertex from
$\mathcal{L}_2$ in Fig.~\ref{fig:cancel}(b), conforms to the naive
$\eps$ power counting, i.e., is $O(\eps)$. 

Similarly, the diagram in Fig.~\ref{fig:cancel}(c) representing the
boson self-energy with one $\mu$ insertion gives
\begin{align}
  -i&\Pi_\mathrm{c}(p)
  = 2\mu\,g^2\int\!\frac{dk}{(2\pi)^{d+1}}\\
  &\qquad\times\left[G_{11}\!\left(k+\frac p2\right)^2
  -G_{12}\!\left(k+\frac p2\right)^2\right]
  G_{22}\!\left(k-\frac p2\right), \notag
\end{align}
which also contains a $1/\eps$ singularity, and hence is
$O(\eps)$ instead of the naive $O(\eps^2)$.  The leading part of
this diagram is 
\begin{equation}
 \Pi_\mathrm{c}(p) = -2\mu\,g^2\int_\k \frac1{4\ek^{\,2}} + \cdots 
  = -2\mu + O(\eps^2),
\end{equation}
which is canceled out exactly by adding the second vertex from
$\mathcal{L}_2$.  Then the sum of Figs.~\ref{fig:cancel}(c) and
\ref{fig:cancel}(d) is again $O(\eps^2)$, consistent with the
naive power counting. 

The $\varphi$ tadpole diagram with one $\mu$ insertion
in Fig.~\ref{fig:cancel}(e) is
\begin{equation}
 \begin{split}
  \Xi_\mathrm{e} &= -\mu g\int\!\frac{dk}{(2\pi)^{d+1}}
  \left[G_{11}(k)-G_{22}(k)\right]G_{21}(k)\\
  &= ig\mu\phi_0\int_\k \frac{\ek}{2E_\k^{\,3}}
  =\frac{ig\mu}{\eps\phi_0}\left(\frac{m\phi_0}{2\pi}\right)^2
  +O(\eps^{3/2}),
 \end{split}
\end{equation}
which is $O(\eps^{1/2})$ instead of the naive $O(\eps^{3/2})$.  
This tadpole diagram should be canceled by other tadpole diagrams of
order $O(\eps^{1/2})$ in Fig.~\ref{fig:cancel}(f), 
\begin{equation}
 \begin{split}
  \Xi_\mathrm{f} 
  &= g\int\!\frac{dk}{(2\pi)^{d+1}}G_{21}(k)-i\frac{g\phi_0}{c_0}\\
  &= -\frac{ig}2\left(\frac{m\phi_0}{2\pi}\right)^2-i\frac{g\phi_0}{c_0} 
  +O(\eps^{3/2}).
 \end{split}
\end{equation}
The condition of cancellation, $\Xi_\mathrm{e}+\Xi_\mathrm{f}=0$, gives
the gap equation that determines $\phi_0(\mu)$ to the leading order in
$\eps$.  The solution to the gap equation is
\begin{equation}
 \phi_0= \frac{2\mu}\eps
  -\frac2{c_0}\left(\frac{2\pi}m\right)^2 + O(\eps).
\end{equation}
When $c_0<0$, this solution can be written in terms of the binding
energy $\eb$ as $\phi_0=(2\mu+\eb)/\eps$.  The previously made
assumption $\mu/\phi_0=O(\eps)$ is indeed correct.  As we will see
later, the cancellation of the tadpole diagrams is automatically
achieved by the minimization of the effective potential. 

Finally, the one-loop vacuum diagram with one $\mu$ insertion in
Fig.~\ref{fig:cancel}(g) also contains the $1/\eps$ singularity as 
\begin{equation}
 \begin{split}
  &i\mu\int\!\frac{dk}{(2\pi)^{d+1}}\left[G_{11}(k)-G_{22}(k)\right]\\
  &\ =\mu\int_\k \frac{\ek}{E_\k}
  =-\frac\mu\eps\left(\frac{m\phi_0}{2\pi}\right)^2 +O(\eps).
 \end{split}
\end{equation}
The leading part of this diagram is $O(1)$ instead of naive
$O(\eps)$ and cannot be canceled by any other diagrams.
Therefore, Fig.~\ref{fig:cancel}(g) $\sim O(1)$ is the only exception of
our naive power counting rule of $\eps$. 

Thus, we can now develop a diagrammatic technique for the Fermi gas near 
the unitarity limit where $g^2/|c_0|\sim\mu$ in terms of the systematic 
expansion over $\eps=4-d$.  The power counting rule of $\eps$ is
summarized as follows: 
\begin{enumerate}
 \item We consider $\mu/\phi_0\sim\eps$ and regard $\phi_0$ as $O(1)$.
 \item For any Green's function, we write down all Feynman diagrams
       according to the Feynman rules in Fig.~\ref{fig:feynman_rules}
       using the propagators from $\mathcal{L}_0$ and the vertices from 
       $\mathcal{L}_1$. 
 \item If there is any subdiagram of the type in Fig.~\ref{fig:cancel}(a) 
       or \ref{fig:cancel}(c), we add the same Feynman diagram where the
       subdiagram is replaced by the vertex from $\mathcal{L}_2$,
       Fig.~\ref{fig:cancel}(b) or \ref{fig:cancel}(d).
 \item The power of $\eps$ for the given Feynman diagram will be
       $O(\eps^{N_g/2+N_\mu})$, where $N_g$ is the number of
       couplings $g$ and $N_\mu$ is the number of chemical potential
       insertions.
 \item The only exception is the diagram in Fig.~\ref{fig:cancel}(g) which
       is $O(1)$.
\end{enumerate}
We note that the sum of all tadpole diagrams in Figs.~\ref{fig:cancel}(e)
and \ref{fig:cancel}(f) vanishes if $\phi_0$ is the solution of the gap 
equation.

\subsection{Effective potential to leading and next-to-leading orders}

\begin{figure}[tp]
 \includegraphics[width=0.40\textwidth,clip]{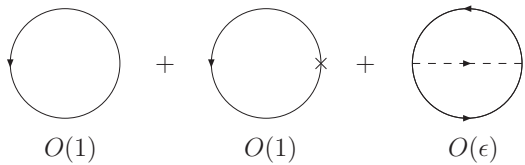}
 \caption{Vacuum diagrams contributing to the effective potential up to 
 the next-to-leading order in $\eps$. The second diagram is $O(1)$
 instead of naive $O(\eps)$ because of the $1/\eps$
 singularity.  \label{fig:potential}}
\end{figure}

We now perform calculations to the leading and next-to-leading orders in
$\eps$, employing the Feynman rules and the $\eps$ power counting
that have just been developed.  To find the dependence of $\phi_0$ on
$\mu$, we use the effective potential method~\cite{Peskin:1995ev} in
which this dependence follows from the minimization of the effective 
potential $V_\mathrm{eff}(\phi_0)$.  
The effective potential at the tree level is given by
\begin{equation}\label{eq:V_0}
 V_0(\phi_0) = \frac{\phi_0^{\,2}}{c_0}. 
\end{equation}
Up to the next-to-leading order, the effective potential receives 
contributions from three vacuum diagrams drawn in
Fig.~\ref{fig:potential}: fermion loops with and without a $\mu$
insertion and a fermion loop with one boson exchange.  

The contribution from two one-loop diagrams is $O(1)$ and given by
\begin{equation}
 \begin{split}
  V_1(\phi_0) &= i\int\!\frac{dp}{(2\pi)^{d+1}}
  \Tr\left[\ln G^{-1}(p) + \mu\sigma_3G(p)\right] \\
  &= -\int_\p \left(E_\p-\mu\frac\ep{E_\p}\right).
 \end{split}
\end{equation}
Using the formula in Eq.~(\ref{eq:dimreg}), we can perform the
integration over $\p$ to find
\begin{align}
  & V_1(\phi_0)= \\
  & -\frac{\left(\frac{m\phi_0}{2\pi}\right)^{\frac d2}}
  {2\Gamma(\frac d2)}\left[\frac{\Gamma(\frac d4)
  \Gamma(-\frac12-\frac d4)}{\Gamma(-\frac12)}\phi_0
  -\frac{\Gamma(\frac d4+\frac12)\Gamma(-\frac d4)}{\Gamma(\frac12)}
  \mu\right]. \notag
\end{align}
Substituting $d=4-\eps$ and expanding in terms of $\eps$ up to
$O(\eps)$, this contribution becomes
\begin{align}\label{eq:V_1}
  V_1(\phi_0) 
  &= \frac{\phi_0}3\left[1+\frac{7-3(\gamma+\ln2)}6\eps\right] 
  \left(\frac{m\phi_0}{2\pi}\right)^{d/2}\\
  &-\frac\mu\eps\left[1+\frac{1-2(\gamma-\ln2)}4\eps\right] 
  \left(\frac{m\phi_0}{2\pi}\right)^{d/2}\!+O(\eps^2), \notag
\end{align}
where $\gamma\approx0.57722$ is the Euler-Mascheroni constant.

The contribution of the two-loop diagram is $O(\eps)$ and given by
\begin{align}
  V_2(\phi_0) 
  & = g^2\int\!\frac{dp\,dq}{(2\pi)^{2d+2}}
  \Tr\left[G(p)\sigma_+G(q)\sigma_-\right]D(p-q) \notag\\
  & = g^2\int\!\frac{dp\,dq}{(2\pi)^{2d+2}}\,
  G_{11}(p)G_{22}(q)D(p-q).
\end{align}
Performing the integrations over $p_0$ and $q_0$, we obtain
\begin{equation}
 V_2(\phi_0) = -\frac{g^2}4\! \int_{\p\q} \frac{(E_\p-\ep)(E_\q-\eq)}
 {E_\p E_\q (E_\p+E_\q+\varepsilon_{\p-\q}/2)}.
\end{equation}
Since this is a next-to-leading-order diagram and the integrals converge
at $d=4$, we can evaluate the integrations at $d=4$.  Changing the
integration variables to $x=\ep/\phi_0$, $y=\eq/\phi_0$, and
$\cos\theta=\hat\p\cdot\hat\q$, the integral can be transformed into
\begin{align}\label{eq:V_2}
  V_2(\phi_0) &= 
  -\eps\left(\frac{m\phi_0}{2\pi}\right)^{d/2}\frac{\phi_0}\pi
  \int_0^\infty\!dx\int_0^\infty\!dy\int_0^\pi\!d\theta\\ 
  &\qquad \times xy\sin^2\theta \, \notag
  \frac{[f(x)-x][f(y)-y]}{f(x)f(y)\left[g(x,y)-\sqrt{xy}\cos\theta\right]},
\end{align}
with $f(x)=\sqrt{x^2+1}$ and $g(x,y)=f(x)+f(y)+\frac12(x+y)$. The
integration over $\theta$ can be performed analytically to lead to
\begin{equation}
  V_2(\phi_0)=-C\eps\left(\frac{m\phi_0}{2\pi}\right)^{d/2}\phi_0,
\end{equation}
where the constant $C$ is given by a two-dimensional integral
\begin{equation}
 \begin{split}
  C &= \int_0^\infty\!dx\int_0^\infty\!dy\, 
 \frac{[f(x)-x][f(y)-y]}{f(x)f(y)} \\
 &\qquad\qquad\qquad \times\left[g(x,y) - \sqrt{g^2(x,y)-xy}\right].
 \end{split}
\end{equation}
A numerical integration over $x$ and $y$ gives
\begin{equation}\label{eq:C}
  C\approx 0.14424.  
\end{equation}

Now, gathering up Eqs.~(\ref{eq:V_0}), (\ref{eq:V_1}), and
(\ref{eq:V_2}), we obtain the effective potential up to the
next-to-leading order in $\eps$,
\begin{widetext}
 \begin{equation}\label{eq:Veff}
 \begin{split}
  V_\mathrm{eff}(\phi_0)&=V_0(\phi_0)+V_1(\phi_0)+V_2(\phi_0)\\
  &= \frac{\phi_0^{\,2}}{c_0} + \left[\frac{\phi_0}3
  \left\{1+\frac{7-3(\gamma+\ln2)}6\eps-3C\eps\right\}
  -\frac\mu\eps\left\{1+\frac{1-2(\gamma-\ln2)}4\eps\right\}
  \right]\left(\frac{m\phi_0}{2\pi}\right)^{d/2}\!+O(\eps^2).
 \end{split} 
 \end{equation}
\end{widetext}
The condition that $\phi_0$ minimizes the effective potential gives the
gap equation; $\d V_\mathrm{eff}/\d\phi_0=0$.  The solution to this
equation is
\begin{align}
  \phi_0 &= \frac{2\mu}\eps\left[1+\left(3C-1+\ln2\right)\eps\right] \\ 
  &\quad -\frac{2\phi_0^{\,\eps/2}}{c_0}
  \left(\frac{2\pi}{m}\right)^{d/2}
  \left[1+\left(3C-1+\frac{\gamma+\ln2}2\right)\eps\right]. \notag
\end{align}
Using the relation of $c_0$ with the binding energy of boson $\eb$ in 
Eq.~(\ref{eq:eb}), we can rewrite the solution of the gap equation in
terms of $\eb$ up to the next-to-leading order in $\eps$ as
\begin{align}\label{eq:phi_0}
  \phi_0 &= \frac{2\mu}\eps\left[1+\left(3C-1+\ln2\right)\eps\right] \\ 
  &\quad +\frac\eb\eps 
  \left[1+\left(3C-\frac12+\ln2
  -\frac12\ln\frac\eb{\phi_0}\right)\eps\right]. \notag
\end{align}
We note that the leading term in Eq.~(\ref{eq:phi_0}) could be
reproduced using the mean field approximation, but the $O(\eps)$
corrections are not.  The $O(\eps)$ correction proportional to the
constant $C$ is a result of the summation of fluctuations around the
classical solution and is beyond the mean field approximation.

\subsection{Thermodynamic quantities near unitarity}
The value of the effective potential $V_\mathrm{eff}$ at its minimum
determines the pressure $P=-V_\mathrm{eff}(\phi_0)$ at a given chemical
potential $\mu$ and a given binding energy of boson $\eb$.  Substituting
the dependence of $\phi_0$ on $\mu$ and $\eb$ in Eq.~(\ref{eq:phi_0}),
we obtain the pressure as 
\begin{equation}\label{eq:pressure}
 P = \frac{\phi_0}6
  \left[ 1+ \left(\frac{17}{12}{-}3C{-}\frac{\gamma{+}\ln 2}2\right)\eps
   -\frac{3\eb}{4\phi_0}\right]\left(\frac{m\phi_0}{2\pi}\right)^{d/2}.
\end{equation}
The fermion number density $N$ is determined by differentiating the
pressure in terms of $\mu$ as
\begin{equation}\label{eq:N}
 N = \frac{\d P}{\d\mu} 
  = \frac1\eps \left[1+\frac{1-2\gamma+2\ln2}4\eps\right]
  \left(\frac{m\phi_0}{2\pi}\right)^{d/2}.
\end{equation}
Then the Fermi energy of a $d$-dimensional free gas with the same number
density is
\begin{equation}\label{eq:eF}
  \eF = \frac{2\pi}m 
   \left[ \frac12\Gamma\left(\frac d2+1\right) N \right]^{2/d}
   = \frac{\phi_0}{\eps^{2/d}}\left(1-\frac{1-\ln2}4\eps\right).
\end{equation}
The nontrivial power of $\eps$ appears because we have raised
$N\sim\eps^{-1}$ to the power of $2/d$.

From Eqs.~(\ref{eq:phi_0}) and (\ref{eq:eF}), we can determine the ratio
of the chemical potential and the Fermi energy $\mu/\eF$ near the
unitary limit as 
\begin{align}\label{eq:mu}
  \frac\mu\eF &=\frac{\eps^{3/2}}2 \notag
  \exp\!\left(\frac{\eps\ln\eps}{8-2\eps}\right) 
  \left[1 - \left(3C -\frac54 (1-\ln2)\right)\eps\right]\\
  &\qquad -\frac{\eb}{2\eF}
  \left[1+\frac\eps2+\frac{\eps\ln\eps}4-\frac\eps2\ln\frac\eb\eF\right].
\end{align}
The $\ln\eps$ term in the second line originates from the
substitution of $\phi_0=\eps^{1/2}\eF$ into the $\ln\eb/\phi_0$ term in
Eq.~(\ref{eq:phi_0}).  We note again that if one is interested in the
extrapolation to $d=3$, one can use Eq.~(\ref{eq:ebeF}) to relate
$\eb/\eF$ to $a\kF$.  The first term in Eq.~(\ref{eq:mu}) gives the
universal parameter of the unitary Fermi gas
$\xi\equiv\left.\mu/\eF\right|_{\eb=0}$,
\begin{equation}\label{eq:xi}
 \begin{split}
  \xi &= \frac{\eps^{3/2}}2 \left[ 1 + \frac{\eps\ln\eps}8 
  - \left(3C -\frac54 (1-\ln2)\right)\eps\right]\\
  &= \frac12\eps^{3/2} + \frac1{16}\eps^{5/2}\ln\eps
  - 0.0246\,\eps^{5/2} + \cdots. 
 \end{split}
\end{equation}
Here we have used the numerical value for $C\approx0.14424$ in 
Eq.~(\ref{eq:C}). The smallness of the coefficient in front of
$\eps^{5/2}$ is a result of a cancellation between the two-loop 
correction and the subleading terms from the expansion of the one-loop
diagrams around $d=4$. 

Using Eqs.~(\ref{eq:pressure}), (\ref{eq:N}), and (\ref{eq:eF}), the
pressure near the unitarity limit normalized by $\eF N$ is given by
\begin{equation}\label{eq:P}
 \frac{P}{\eF N} = \frac2{d+2}\xi-\frac{\eb}{8\eF}\eps.
\end{equation}
Then the energy density $E=\mu N-P$ can be calculated from
Eqs.~(\ref{eq:mu}) and (\ref{eq:P}) as
\begin{equation}\label{eq:E}
 \frac{E}{\eF N} = \frac d{d+2}\xi
  -\frac{\eb}{2\eF}
  \left[1+\frac\eps4+\frac{\eps\ln\eps}4-\frac\eps2\ln\frac\eb\eF\right].
\end{equation}
The pressure and energy density in the unitarity limit are obtained 
from $\xi$ via the universal relations depending only on the
dimensionality of space.

\subsection{Quasiparticle spectrum}
\begin{figure}[tp]
 \includegraphics[width=0.5\textwidth,clip]{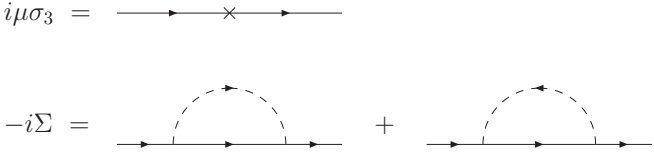}
 \caption{Corrections to the fermion self-energy of order $O(\eps)$;
 a $\mu$ insertion and one-loop diagrams. \label{fig:self_energy}}
\end{figure}

The $\eps$ expansion we have developed is also useful for the 
calculations of physical
observables other than the thermodynamic quantities.  Here we shall look 
at the dispersion relation of fermion quasiparticles.  To the leading
order in $\eps$, the dispersion relation is given by
$\omega_\mathrm{F}(\p)=E_\p=\sqrt{\ep^{\,2}+\phi_0^{\,2}}$, which has a
minimum at zero momentum $\p=\bm0$ with the energy gap equal to 
$\Delta=\phi_0=(2\mu+\eb)/\eps$.  The next-to-leading-order corrections
to the dispersion relation come from three sources: from the correction
of $\phi_0$ in Eq.~(\ref{eq:phi_0}), from a $\mu$ insertion to the
fermion propagator, and from the one-loop self-energy diagrams
$-i\Sigma(p)$ depicted in Fig.~\ref{fig:self_energy}. 

Using the Feynman rules, the one-loop diagram of the fermion self-energy
in Fig.~\ref{fig:self_energy} is given by 
\begin{equation}\label{eq:self-energy}
 \begin{split}
  -i\Sigma(p) = g^2\int\!\frac{dk}{(2\pi)^{d+1}}\,
 &\left[\sigma_+G(k)\sigma_-D(p-k)\right.\\
 &\left.+\sigma_-G(k)\sigma_+D(k-p)\right].
 \end{split}
\end{equation}
There are corrections only to the diagonal elements of the self-energy
and each element is evaluated as
\begin{equation}\label{eq:Sigma11}
 \begin{split}
  \Sigma_{11}(p) &= ig^2\int\!\frac{dk}{(2\pi)^{d+1}}\,G_{22}(k)D(p-k)\\
  &=-\frac{g^2}2\int_\k \frac{E_\k-\ek}
  {E_\k (E_\k + \varepsilon_{\k-\p}/2-p_0-i\delta)},
 \end{split}
\end{equation}
and
\begin{equation}
 \begin{split}
  \Sigma_{22}(p) &= ig^2\int\!\frac{dk}{(2\pi)^{d+1}}\,G_{11}(k)D(k-p)\\
  &=\frac{g^2}2\int_\k \frac{E_\k-\ek}
  {E_\k (E_\k + \varepsilon_{\k-\p}/2+p_0-i\delta)}.
 \end{split}
\end{equation}

The dispersion relation of the fermion quasiparticle
$\omega_\mathrm{F}(\p)$ is obtained as a pole of the fermion propagator 
$\det[G^{-1}(\omega,\p)+\mu\sigma_3-\Sigma(\omega,\p)]=0$, which reduces
to the following equation: 
\begin{equation}\label{eq:dispersion}
 \begin{vmatrix}
   \omega-\ep+\mu-\Sigma_{11}(\omega,\p)&\phi_0\\
   \phi_0&\omega+\ep-\mu-\Sigma_{22}(\omega,\p)
 \end{vmatrix}=0.
\end{equation}
To find the $O(\eps)$ correction to the dispersion relation, we only
have to evaluate the self-energy $\Sigma(\omega,\p)$ with $\omega$
given by the leading order solution $\omega=\Ep$.  Denoting
$\Sigma_{11}(\Ep,\p)$ and $\Sigma_{22}(\Ep,\p)$ simply by $\Sigma_{11}$
and $\Sigma_{22}$ and solving Eq.~(\ref{eq:dispersion}) in terms of
$\omega$, we obtain the dispersion relation of the fermion quasiparticle
as
\begin{equation}\label{eq:omega_F}
 \omega_\mathrm{F}(\p)=\Ep+\frac{\Sigma_{11}+\Sigma_{22}}2
  +\frac{\Sigma_{11}-\Sigma_{22}-2\mu}{2\Ep}\ep +O(\eps^2). 
\end{equation}

In order to find the energy gap of the fermion quasiparticle, since the
minimum of the dispersion relation will appear at small momentum
$\ep\sim\mu$, we can expand $\Sigma(\Ep,\p)$ around zero momentum $\p=0$
as
$\Sigma(\Ep,\p)=\Sigma^0(\phi_0,\bm0)+\Sigma'(\phi_0,\bm0)\,\ep/\phi_0$. 
Performing the integration over $\k$ in $\Sigma$ analytically, we have
\begin{equation}
 \begin{split}
  \Sigma_{11}(\phi_0,\p)&=\eps\left(2-8\ln3+8\ln2\right)\phi_0\\
  &\qquad+\eps\left(-\frac83+8\ln3-8\ln2\right)\ep,
 \end{split}
\end{equation}
and
\begin{equation}
 \begin{split}
  \Sigma_{22}(\phi_0,\p)&=\eps\left(-2-8\ln3+16\ln2\right)\phi_0\\
  &\qquad+\eps\left(-\frac73-8\ln3+16\ln2\right)\ep.
 \end{split}
\end{equation}
Introducing these expressions into Eq.~(\ref{eq:omega_F}), we find 
the fermion dispersion relation around its minimum has the following
form:
\begin{equation}\label{eq:omega}
 \omega_\mathrm{F}(\p)
  \simeq\Delta+\frac{(\ep-\varepsilon_0)^2}{2\phi_0} 
  \simeq\sqrt{(\ep-\varepsilon_0)^2+\Delta^2}. 
\end{equation}

\begin{figure}[tp]
 \includegraphics[clip]{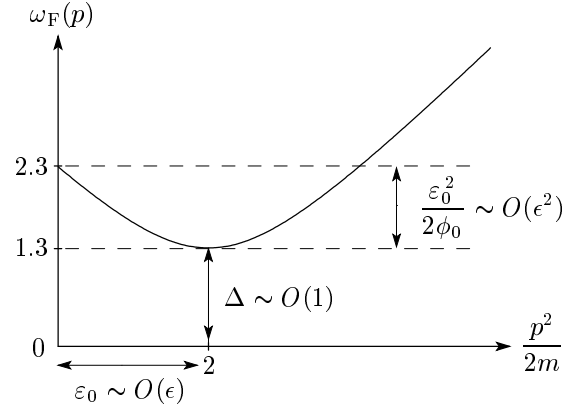}
 \caption{Illustration of the dispersion relation of the fermion
 quasiparticle in the unitarity limit from the $\eps$ expansion. 
 $\omega_\mathrm{F}(\p)=\sqrt{(\ep-\varepsilon_0)^2+\Delta^2}$ in
 Eq.~(\ref{eq:omega}) is plotted as a function of
 $\ep=\p^2/2m$.  Values on both axes are extrapolated results to
 $\eps=1$ in units of the chemical potential $\mu$.
 \label{fig:dispersion}}
\end{figure}

Here $\Delta$ is the energy gap of the fermion quasiparticle, which is
given by 
\begin{equation}\label{eq:gap}
 \begin{split}
  \Delta &= \phi_0 + \frac{\Sigma_{11}^0+\Sigma_{22}^0}2 \\
  &= \left[1-\left(8\ln3-12\ln2\right)\eps+O(\eps^2)\right]\phi_0.
 \end{split}
\end{equation}
The minimum of the dispersion curve is located at a nonzero value of
momentum, $|\p|=(2m\varepsilon_0)^{1/2}$, where
\begin{equation}
 \begin{split}
  \varepsilon_0 &= \mu + \frac{\Sigma_{22}^0-\Sigma_{11}^0}2 
  - \frac{\Sigma_{11}'+\Sigma_{22}'}2 \\
  &= \mu + \frac{\eps\phi_0}2 +O(\eps^2).
 \end{split}
\end{equation}
Then, introducing the solution of the gap equation (\ref{eq:phi_0}), 
the energy gap $\Delta$ as a function of the chemical potential and the
binding energy is given by
\begin{align}\label{eq:gap_mu}
 \Delta &= \frac{2\mu}\eps\left[1+(3C-1-8\ln3+13\ln2)\eps\right] \notag\\ 
 &\quad +\frac\eb\eps\left[1+\left(3C-\frac12-8\ln3+13\ln2
 -\frac12\ln\frac\eb{\Delta}\right)\eps\right] \notag\\
 &= \frac{2\mu}\eps\left(1-0.345\,\eps\right) \\ 
 &\quad + \frac\eb\eps\left[1+0.155\,\eps-\frac{\eps\ln\eps}2
 +\frac\eps2\ln\!\left(1+\frac{2\mu}\eb\right)\right], \notag 
\end{align}
while $\varepsilon_0$ is given by
\begin{equation}\label{eq:e_0}
 \varepsilon_0=2\mu+\frac\eb2.
\end{equation}
Note the difference with the mean field approximation, in which
$\varepsilon_0=\mu$.  

When $\varepsilon_0$ is positive, the fermion dispersion curve has its
minimum at nonzero value of momentum as in the BCS limit, while the
minimum is located at zero momentum when $\varepsilon_0$ is negative as
in the BEC limit.  We find the former $(\varepsilon_0=2\mu>0)$ is the
case in the unitarity limit.  The dispersion curve of the fermion
quasiparticle $\omega_\mathrm{F}(\p)$ in the unitarity limit $\eb=0$ is
illustrated in Fig~\ref{fig:dispersion}.  In particular, the difference
between the fermion quasiparticle energy at zero momentum and at its
minimum is given by 
$\varepsilon_0^{\,2}/2\phi_0=\eps\mu\sim O(\eps^2)$.

\subsection{Location of the splitting point}
Let us consider the situation where the binding energy $\eb$ is
increasing from zero while the number density $N$ is kept fixed.  Then 
$\Delta>0$ is held fixed but $\mu$ is changing.  Since to the leading 
order in $\eps$, the chemical potential as a function of the energy gap 
is given by $2\mu=\eps\Delta-\eb$, the location of the minimum of the 
dispersion curve can be written as 
\begin{equation}\label{eq:e0}
 \varepsilon_0=\eps\Delta-\frac\eb2.
\end{equation}
Therefore, $\varepsilon_0$ decreases as $\eb$ increases. 
When the binding energy reaches $\eb=2\eps\Delta$, the minimum of the
dispersion curve sits exactly at zero momentum $\p=\bm0$.  This point is
referred to as a \textit{splitting point}~\cite{son05}.
We find the splitting point is located on the BEC side of the unitarity
limit where the binding energy is positive $\eb=2\eps\Delta>0$ and the 
chemical potential is negative $2\mu=-\eps\Delta$.
This splitting point will play an important role to determine the phase 
structure of the polarized Fermi gas in the unitary regime as we will
study in Sec.~\ref{sec:polarization}.

\subsection{Extrapolation to $\epsilon=1$}
Finally, we discuss the extrapolation of the series expansion to the
physical case in three spatial dimensions.  
Although the formalism is based on the smallness of $\eps$, we see
that the next-to-leading-order corrections are reasonably small compared
to the leading terms even at $\eps=1$.  If we naively use only the
leading and next-to-leading-order results for $\xi$ in Eq.~(\ref{eq:xi}),
$\Delta$ in Eq.~(\ref{eq:gap_mu}), and $\varepsilon_0$ in
Eq.~(\ref{eq:e_0}) in the unitarity limit, their extrapolations to
$\eps=1$ give for three spatial dimensions
\begin{equation}\label{eq:extrapolation}
 \xi \approx 0.475, \qquad \frac{\Delta}{\mu}\approx 1.31, 
  \qquad \frac{\varepsilon_0}{\mu}\approx 2. 
\end{equation}
They are reasonably close to the results of recent Monte Carlo 
simulations, which yield $\xi\approx0.42$, $\Delta/\mu\approx1.2$, 
and $\varepsilon_0/\mu\approx1.9$~\cite{Carlson:2005kg}.  
They are also consistent with recent experimental measurements of $\xi$,
where $\xi=0.51\pm0.04$~\cite{Thomas05} and
$\xi=0.46\pm0.05$~\cite{Hulet-polarized}.  These
agreements can be taken as a strong indication that the $\eps$
expansion is useful even at $\eps=1$. 

Our $\eps$ expansion predicts the behavior of the thermodynamic
quantities near the unitarity limit.  From Eqs.~(\ref{eq:P}),
(\ref{eq:E}), and (\ref{eq:mu}), the extrapolations to the three spatial 
dimensions $\eps=1$ lead to
\begin{align}
 \frac{P}{\eF N} &\approx \frac2{d+2}\,\xi -\frac18\frac\eb\eF, \\
 \frac{E}{\eF N} &\approx \frac{d}{d+2}\,\xi 
 - \frac14\frac\eb\eF\left(\frac52-\ln\frac\eb\eF\right), 
 \phantom{\frac{\frac\int\int}{\frac\int\int}}\hspace{-4.5mm} \\ 
 \frac{\mu}{\eF} &\approx \xi 
 - \frac14\frac\eb\eF\left(3-\ln\frac\eb\eF\right).
\end{align}
The pressure, energy density, and chemical potential at fixed density
are decreasing functions of the binding energy $\eb$ with the slope shown
above.  
In Sec.~\ref{sec:matching}, we will make a discussion to improve the
extrapolation of the series expansion by imposing the exact result at
$d=2$ as a boundary condition. 

We can also try to determine the location of the splitting point.  At
this point, the deviation from the unitarity point measured by the
binding energy per Fermi energy is given by 
\begin{equation}
  \frac\eb\eF = \frac2{\eps^{2/d-1}}[1+O(\eps)] \approx 2 \qquad
  \text{at}\ \  d=3. 
\end{equation}
Comparing with Eq.~(\ref{eq:ebeF}), one finds $a\kF\approx 1$ at the
splitting point.  Since this result is known only to the leading order
in $\eps$, one must be cautious with the numerical value.  However,
certain qualitative features are probably correct: the splitting point
is located on the BEC side of the unitarity limit ($a>0$) and the
chemical potential $\mu/\eF\approx-0.5$ is negative at this point.

\section{Phase structure of polarized Fermi gas \label{sec:polarization}}

\subsection{Effective potential at finite superfluid velocity}
Here we apply the $\eps$ expansion developed in the previous section to
the unitary Fermi gas with unequal densities for the two fermion
components.  We put special emphasis on investigating the phase
structure of polarized Fermi gas in the unitary regime, which has a
direct relation with the recent measurements in atomic
traps~\cite{Ketterle-polarized,Hulet-polarized,Ketterle-polarized2,Ketterle-polarized3}. 
For this purpose, we generalize our formalism to take into account
the possibility of a spatially varying condensate where 
$\<\phi(x)\>=e^{2im\vs\cdot\bm x}\phi_0$ with $\vs$ being the superfluid
velocity.  The factor $e^{2im\vs\cdot\bm x}$ in front of $\phi_0$ in the
Lagrangian density can be absorbed by making the Galilean transformation
on the fermion field as $\psi_\sigma(x)\to e^{im\vs\cdot\bm
x}\psi_\sigma(x)$ and the boson field as 
$\varphi(x)\to e^{2im\vs\cdot\bm x}\varphi(x)$. Accordingly, 
the fermion propagator in Eq.~(\ref{eq:G}) is modified as 
\begin{equation}\label{eq:G_vs}
 \begin{split}
  G(p_0,\p) &= \frac1{(p_0+H-\p\cdot\vs)^2
  -(\ep+\varepsilon_{m\vs})^2-\phi_0^{\,2}+i\delta}\\ &\quad \times
  \begin{pmatrix}
   p_0 + H + \varepsilon_{\p-m\vs} & -\phi_0 \\
   -\phi_0 & p_0+ H - \varepsilon_{\p+m\vs}
  \end{pmatrix},
 \end{split}
\end{equation}
while the boson propagator in Eq.~(\ref{eq:D}) is modified as
\begin{equation}
 D(p_0,\p) = \left(p_0 - \frac{\varepsilon_{\p+2m\vs}}2 
	      + i\delta\right)^{-1}.
\end{equation}

The value of the superfluid velocity $\vs$ is determined by minimizing
the $\vs$ dependent part of the effective potential $V_H(\vs)$.  As far
as the polarization $H$ is sufficiently small compared to the energy gap
$\Delta$ of the fermion quasiparticle, $V_H(\vs)$ to the leading order
in $\eps$ is given by the fermion one-loop diagram with the propagator
in Eq.~(\ref{eq:G_vs}) as follows: 
\begin{equation}
 V_{H<\Delta}(\vs)=-\int_\p \sqrt{\left(\ep+\ev\right)^2+\phi_0^{\,2}}.
\end{equation}
Since it will turn out that $\ev$ is $O(\eps^{11})$, we can expand the 
integrand in terms of $\ev/\phi_0$ to lead to 
\begin{equation}
 V_{H<\Delta}(\vs)=-\ev\int_\p \frac{\ep}{E_\p}
  \simeq\frac{\ev}\eps\left(\frac{m\phi_0}{2\pi}\right)^2.
\end{equation}
Using the result on the fermion number density $N$ in Eq.~(\ref{eq:N}),
this part can be rewritten as $V_{H<\Delta}(\vs)=N\ev$, which represents
the energy cost due to the presence of the superfluid flow. 

If $H-\p\cdot\vs$ reaches the bottom of the fermion quasiparticle
spectrum $\Delta+(\ep+\ev-\varepsilon_0)^2/2\phi_0$, $V_H(\vs)$ receives 
an additional contribution from the filled fermion quasiparticles. 
Since $\ev\sim\eps^{11}$ is small, we can neglect it in the
quasiparticle energy.  Then the $\vs$ dependent part of the effective  
potential is given by
\begin{equation}
 V_H(\vs)=N\ev-\int_\p
  \left[H-\p\cdot\vs-\omega_\mathrm{F}(\p)\right]_>,
\end{equation}
where we have introduced a notation $(x)^y_>=x^y\,\theta(x)$ and
$\omega_\mathrm{F}(\p)$ is the fermion quasiparticle spectrum
$\omega_\mathrm{F}(\p)=\Delta+(\ep-\varepsilon_0)^2/2\phi_0$ derived in 
Eq.~(\ref{eq:omega}). The effective potential $V_H(\vs)$ to the leading
order in $\eps$ has the same form as that studied based on the effective
field theory in~\cite{son05}.

\subsection{Critical polarizations}
Now we evaluate the effective potential $V_H(\vs)$ at $d=4$ as a
function of $\vs$ and $H$.  Changing the integration variables to
$z=\ep/\phi_0$ and $w=\hat\p\cdot\hat{\bm{v}}_\mathrm{s}$, we have 
\begin{equation}
 \begin{split}
  V_H(\vs)&=\left(\frac{m\phi_0}{2\pi}\right)^2\left[\frac\ev{\eps}
  -\frac2\pi\int_0^\infty\!dz\!\int_{-1}^1\!dw\,z\sqrt{1-w^2}\right.\\ 
  &\left. \times\left(H-\Delta-\frac{(z-z_0)^2}{2}\phi_0
  -2\sqrt{z\phi_0\ev}w\right)_>\right],
 \end{split}
\end{equation}
where $z_0=\varepsilon_0/\phi_0\sim\eps$.
We can approximate $\sqrt{z\phi_0\ev}$ by $\sqrt{z_0\phi_0\ev}$ because
the difference will be $O(\eps^8)$ and negligible compared to itself 
$\sqrt{z_0\phi_0\ev}\sim\eps^6$.  Then the integration over $z$ leads to 
\begin{equation}
 \begin{split}
  V_H(\vs)&=\left(\frac{m\phi_0}{2\pi}\right)^2\phi_0
  \left[\frac\ev{\eps\phi_0}-\frac{32}{3\pi}z_0
  \left(\frac\ev{\phi_0}z_0\right)^{3/4} \right.\\
  &\left. \times\int_{-1}^1dw\,\sqrt{1-w^2}
  \biggl(\frac{H-\Delta}{2\sqrt{\ev{\phi_0}z_0}}-w\biggr)^{3/2}_>\right].
 \end{split}
\end{equation}
By introducing the dimensionless variables $x$ and $h$ as
\begin{equation}
 \frac\ev{\phi_0}=x^2\left(\frac{32}{3\pi}\eps\right)^4z_0^{\,7},
\end{equation}
and
\begin{equation}
 \frac{H-\Delta}{\phi_0}=2h\left(\frac{32}{3\pi}\eps\right)^2z_0^{\,4},
\end{equation}
we can rewrite the effective potential $V_H(\vs)=V_h(x)$ in the simple 
form as  
\begin{align}
  V_h(x)&=\phi_0\left(\frac{m\phi_0}{2\pi}\right)^2
  \left(\frac{32}{3\pi}\right)^4\eps^3z_0^{\,7}\\
  &\times\left[x^2-x^{3/2}\int_{-1}^1dw\,\sqrt{1-w^2}
  \left(\frac hx-w\right)^{3/2}_>\right]. \notag
\end{align}
Now we see that if the superfluid velocity exists, $\ev\sim\eps^{11}$
and the $\vs$ dependent part of the effective potential is
$O(\eps^{10})$. 

Numerical studies on the effective potential $V_h(x)$ as a
function of $x$ show that there exists a region of $h$, $h_1<h<h_2$,
where $V_h(x)$ has its minimum at finite $x$.  These two critical values
are numerically given by 
\begin{equation}
 h_1=-0.00366\qquad\text{and}\qquad h_2=0.0275.
\end{equation}
Correspondingly, we obtain the critical polarizations normalized by the
energy gap as 
\begin{equation}
  \frac{H_1}\Delta=1-0.0843\,\eps^2
  \left(\frac{\varepsilon_0}\Delta\right)^{4},\\ 
\end{equation}
and
\begin{equation}
  \frac{H_2}\Delta=1+0.634\,\eps^2
  \left(\frac{\varepsilon_0}\Delta\right)^{4}.
\end{equation}
Here we have replaced $\phi_0$ by $\Delta$ because they only differ by
$O(\eps)$ [Eq.~(\ref{eq:gap})]. 
The region for the superfluid phase with the spatially varying
condensate is $H_2-H_1\sim\eps^6$, where the superfluid velocity $\vs$
is finite at the ground state. 

As the polarization increases further $H>H_2$, the superfluid velocity
disappears.  If $H<\omega_\mathrm{F}(\bm0)$, fermion quasiparticles 
which have momentum $\omega_\mathrm{F}(\p)<H$ are filled and hence there
exist two Fermi surfaces, while there is only one Fermi surface for
$H>\omega_\mathrm{F}(\bm0)$ (see Fig.~\ref{fig:dispersion}). 
Therefore, from the quasiparticle spectrum derived in
Eq.~(\ref{eq:omega}), the polarization for the disappearance of the inner
Fermi surface $H_3$ is given by
\begin{equation}
 \frac{H_3}\Delta=\frac{\omega_\mathrm{F}(\bm0)}\Delta
  =1+\frac12\left(\frac{\varepsilon_0}\Delta\right)^2.
\end{equation}
Here the location of minimum in the fermion quasiparticle spectrum
$\varepsilon_0$ is related with the binding energy $\eb$ near the
unitarity limit via $\varepsilon_0=\eps\Delta-\eb/2$
[Eq.~(\ref{eq:e0})].  The critical polarizations $H_1/\Delta$ and
$H_2/\Delta$ as functions of $\eb/\eps\Delta$ are illustrated in
Fig.~{\ref{fig:polarization}}.

\subsection{Phase transition to normal Fermi gas}
Next we turn to the phase transition to the normal Fermi
gas which occurs at $H-\Delta\sim\eps$.  Since the region for the
superfluid phase with the spatially varying condensate is
$H_{1,2}-\Delta\sim\eps^6$, we can neglect the superfluid velocity $\vs$
here.  We can also neglect the possibility to have two Fermi surfaces
where $H_3-\Delta\sim\eps^2$.  Then the contribution of the polarized
quasiparticles to the effective potential is evaluated as
\begin{equation}
 V_H(\bm0) = -\int_\p \left[H-\omega_\mathrm{F}(\p)\right]_> 
  \simeq -\frac{(H-\Delta)_>^2}{2\phi_0}
  \left(\frac{m\phi_0}{2\pi}\right)^{d/2},
\end{equation}
where we have neglected the higher-order corrections due to the shift of
the location of minimum in the dispersion relation
$\varepsilon_0\sim\eps$.  According to the modification of the effective
potential in Eq.~(\ref{eq:Veff}) to $\Veff(\phi_0)+V_H$, the solution of
the gap equation in Eq.~(\ref{eq:phi_0}) becomes
$\phi_0\to\phi_0+\phi_H$, where 
\begin{equation}
\phi_H = -(H-\Delta)_> + \frac{(H-\Delta)^2_>}{2\phi_0}.
\end{equation}
Then the pressure of the polarized Fermi gas in the superfluid state is
given by $P=-\Veff(\phi_0+\phi_H)-V_H$, which results in 
\begin{align}\label{eq:P_H}
 P &= \frac{\phi_0}6 \left[ 1 + \left(\frac{17}{12}{-}3C
 {-}\frac{\gamma{+}\ln 2}2\right)\eps -\frac{3\eb}{4\phi_0} \right]
 \left(\frac{m\phi_0}{2\pi}\right)^{d/2} \notag\\
 &\qquad + O\!\left[\,\eps(H-\Delta)^2_>,\,(H-\Delta)^3_>\,\right].
 \phantom{\frac\int{}}
\end{align}
The polarization dependent part of the pressure $P_H\sim\eps^3$ is small
and negligible in the region considered here $H-\Delta\sim\eps$. 

The pressure of the superfluid state should be compared to that of the
normal state with the same chemical potentials.  Since the phase
transition to the normal state happens at 
$H\sim\phi_0\gg\mu\sim\eps\phi_0$, only one component of fermions exists
in the normal state.  Therefore, the interaction is completely
suppressed in the normal Fermi gas and its pressure $P_\mathrm{n}$ is
simply given by that of a free Fermi gas with a single component as
follows:
\begin{equation}\label{eq:Pn}
 P_\mathrm{n} = \int_\p \left(\mu_\uparrow-\ep\right)_>
  = \frac{(H+\mu)^{\frac d2+1}}{\Gamma\!\left(\frac d2+2\right)}
  \left(\frac{m}{2\pi}\right)^{\frac d2}.
\end{equation}

The phase transition of the superfluid state to the normal state occurs
at $H=\Hc$ where the two pressures coincide $P=P_\mathrm{n}$.  From 
Eqs.~(\ref{eq:P_H}) and (\ref{eq:Pn}), the critical polarization $\Hc$
satisfies the following equation:
\begin{align}\label{eq:Hc}
  \Hc
  &=\left[1+\frac\eb{4\phi_0}-C\eps-\frac{2+\ln2}6\eps\right]\phi_0\\
  &=\left[1+\frac\eb{4\Delta}-C\eps-\frac{2+\ln2}6\eps
  +(8\ln3-12\ln2)\eps\right]\Delta, \notag
\end{align}
where we have substituted the relation between the condensate $\phi_0$ 
and the energy gap $\Delta$ at zero polarization in Eq.~(\ref{eq:gap}). 
Defining a number $\sigma\sim O(1)$ by
\begin{equation}
  \sigma=C+\frac{2+\ln2}6-(8\ln3-12\ln2) = 0.122, 
\end{equation}
the critical polarization normalized by the energy gap at zero
polarization is written as 
\begin{equation}
 \frac{H_\mathrm{c}}\Delta 
  = 1-\eps\sigma+\frac{\eb}{4\Delta} + O(\eps^2). 
\end{equation}
If the binding energy is large enough $\eb/\eps\Delta>4\sigma=0.488$,
the superfluid state remains above $H=\Delta$. 
The superfluid state at $\Delta<H<\Hc$ is referred to as a
\textit{gapless} superfluid state.  The fermion number densities for two
different components are asymmetric there and the fermionic excitation
does not have the energy gap. 

In particular, at the unitarity limit where $\eb/\Delta=0$, the critical
polarization is given by
\begin{equation}
  \left.\frac{H_\mathrm{c}}\Delta\right|_\mathrm{UL}
  =1-\eps\sigma =1-0.122\,\eps. 
\end{equation}
At the splitting point where $\eb/\Delta=2\eps$, it is 
\begin{equation}
  \left.\frac{H_\mathrm{c}}\Delta\right|_\mathrm{SP}
  =1-\eps\sigma+\frac\eps2 =1+0.378\,\eps. 
\end{equation}
The extrapolations to three spatial dimensions $\eps=1$ give the
critical polarizations at the two typical points as
$\Hc/\Delta\,|_\mathrm{UL}=0.878$ and
$\Hc/\Delta\,|_\mathrm{SP}=1.378$.  At unitarity, hence, there is no
gapless superfluid phase due to the competition with the normal phase.
On the other hand, near the splitting point, the normal phase is not
competitive compared to the gapless phases and the gapless superfluid
phases stably exist there.  The phase boundary between the superfluid
and normal phases $\Hc/\Delta$ as a function of $\eb/\eps\Delta$ is
illustrated in Fig.~{\ref{fig:polarization}}.

\subsection{Phase structure near the unitarity limit}

\begin{figure}[tp]
 \includegraphics[clip]{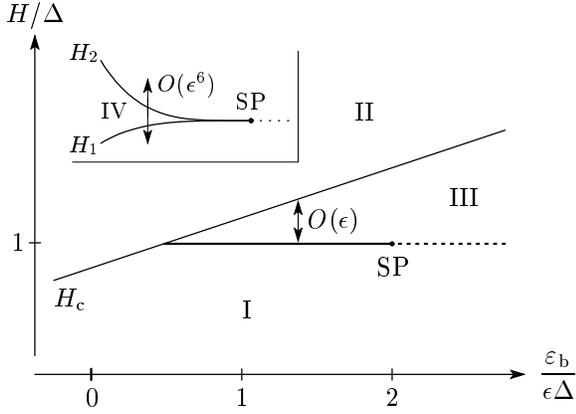}
 \caption{Schematic phase diagram of the polarized Fermi gas near the 
 unitarity limit from the $\eps$ expansion.  The phase diagram can be
 divided into four phases, I: the gapped superfluid phase, II: the
 polarized normal phase, III: the gapless superfluid phase, and IV: the
 phase with the spatially varying condensate.  The inset is the
 magnification of the region around the splitting point (SP).  The phase
 IV appears in the narrow region represented by the thick line between
 the phases I and III.  \label{fig:polarization}}
\end{figure}

The schematic phase diagram of the polarized Fermi gas in the unitary
regime is shown in Fig.~\ref{fig:polarization} in the plane of $H$ and
$\eb/\eps$ for the fixed energy gap $\Delta$ at zero polarization.  The
critical polarization $\Hc/\Delta$ divides the phase diagram into two
regions; the superfluid phase at $H<\Hc$ (I and III) and the normal
phase at $H>\Hc$ (II).  The Fermi gas in the normal phase is fully
polarized near the unitarity limit because $H\gg\mu$.  The phase
transition of the superfluid state to the normal state is of the first
order, because of the discontinuity in the fermion number density which
is $O(\eps^{-1})$ in the superfluid phase while it is $O(1)$ in the
normal phase. 

The superfluid phase can be divided further into two regions; the gapped 
superfluid phase at $H<\Delta$ (I) and the gapless superfluid phase at
$\Delta<H<\Hc$ (III).  On the BEC side of the splitting point (SP) where
$\eb/\eps\Delta>2$, the phase transition from the gapped phase to the 
gapless phase is of the second order because a discontinuity appears in
the second derivative of the pressure 
$\d^2 P/\d H^2\sim\eps\,\theta(H-\Delta)$ [see Eq.~(\ref{eq:P_H})].  
On the other hand, on the BCS side of the splitting point where
$4\sigma<\eb/\eps\Delta<2$, there exists the superfluid phase with the
spatially varying condensate (IV) at $H_1<H<H_2$ between the gapped and
gapless phases.  This phase appears only in the narrow region where
$H_2-H_1\sim\eps^6$.  The phase transitions at $H=H_1,H_2$ are of the
first order. 

In actual experiments using fermionic atoms, unitary Fermi gases are
trapped in optical potentials
$V(r)$~\cite{Ketterle-polarized,Hulet-polarized,Ketterle-polarized2,Ketterle-polarized3}. 
In such cases, the polarization $H$ and the scattering length or $\eb$
are constant over the space, while the effective chemical potential
$\mu-V(r)$ decreases from the center to the peripheral of the trapping
potential.  For the purpose of comparison of our results with the
experiments on the polarized Fermi gases, it is convenient to visualize
the phase diagram in the plane of $H$ and $\mu$ for the fixed $\eb$. 

\begin{figure}[tp]
 \includegraphics[clip]{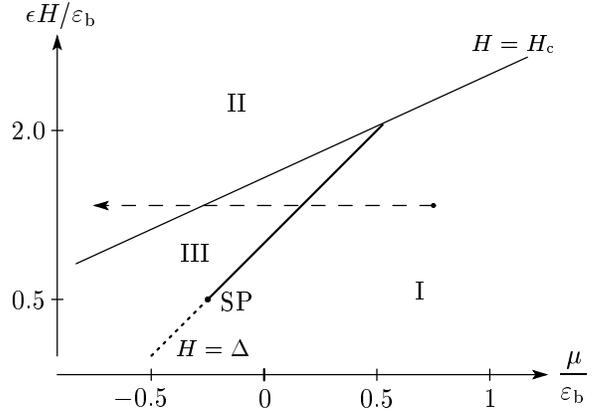}
 \caption{Schematic phase diagram of the polarized Fermi gas near the 
 unitarity limit in the $H$-$\mu$ plane for the fixed binding
 energy $\eb>0$.  The phases I, II, and III are the same as in
 Fig.~\ref{fig:polarization}.  The horizontal dashed line from right to
 left tracks the chemical potentials $(\mu,H)$ in trapped Fermi gases 
 from the center to the peripheral.  \label{fig:polarization2}}
\end{figure}

In Fig.~\ref{fig:polarization2}, $\eps\Hc/\eb$ and $\eps\Delta/\eb$ are
plotted as functions of $\mu/\eb$ for the fixed binding energy $\eb>0$. 
The superfluid phase and the normal phase are separated by the line of
the critical polarization $H=\Hc$, while the gapped and gapless
superfluid phases are separated by the line of the energy gap at zero
polarization $H=\Delta$.  The intersection of the two lines,
$\Hc=\Delta$, is located at 
$(\mu/\eb,\eps H/\eb)=((1-4\sigma)/8\sigma,1/4\sigma)=(0.524,2.05)$,
while the splitting point in the phase diagram is at 
$(\mu/\eb,\eps H/\eb)=(-1/4,1/2)$. 
The horizontal dashed line from right to left tracks the chemical
potential $\mu$ in the trapped Fermi gas from the center to the
peripheral.  If the polarization is small enough compared to the binding
energy $\eps H/\eb<2.05$, there exists the gapless superfluid phase
(III) between the gapped superfluid phase (I) and the normal phase
(II). When the polarization is above the splitting point 
$0.5<\eps H/\eb<2.05$, the phase with the spatially varying condensate 
(thick line between I and III) will appear between the gapped and
gapless superfluid phases. 

Assuming that the above picture remains qualitatively valid in three
spatial dimensions, we thus found that the physics around the splitting
point is the same as argued in Ref.~\cite{son05}.  However, due to the
competition with the normal phase, the gapless phases disappear at some
point, probably before the unitarity is reached.  The point where this
happens can be estimated from our calculations to be
\begin{equation}
  (a\kF)^{-1} \approx \sqrt{2\sigma} = 0.494. 
\end{equation}
The gapless phase with the spatially varying condensate exists in a
finite range $0.494\lesssim(a\kF)^{-1}\lesssim1$.  The hypothesis of
Ref.~\cite{son05} that the region of the gapless phase with the
spatially varying condensate is connected with the FFLO region on the
BCS side is not realized in the $\eps$ expansion.

\section{Expansion around two spatial dimensions \label{sec:2d}}

\subsection{Lagrangian and Feynman rules}
In this section, we formulate the systematic expansion for the unitary
Fermi gas around two spatial dimensions in a similar way as we have done
for the $\eps=4-d$ expansion.  Here we start with the Lagrangian density
given in Eq.~(\ref{eq:L}) limited to the unpolarized Fermi gas in the
unitarity limit where $H=0$ and $1/c_0=0$ as follows:
\begin{equation}
 \begin{split}
  \mathcal{L} &= \Psi^\+\left(i\d_t + \frac{\sigma_3\grad^2}{2m}
  + \mu\sigma_3\right)\Psi \\ &\qquad\qquad\qquad
  + \Psi^\+\sigma_+\Psi\phi + \Psi^\+\sigma_-\Psi\phi^*.
 \end{split}
\end{equation}
Then we expand the field $\phi$ around its vacuum expectation value
$\phi_0$ as 
\begin{equation}\label{eq:coupling_2d}
 \phi=\phi_0 + \bar g\varphi, \qquad 
  \bar g = \left(\frac{2\pi\bar\eps}m\right)^{1/2}
  \left(\frac{m\mu}{2\pi}\right)^{-\bar\eps/4},
\end{equation}
where the effective coupling $\bar g\sim\bar\eps^{1/2}$ in
Eq.~(\ref{eq:g-bar}) was introduced.  The extra factor
$\left(m\mu/2\pi\right)^{-\bar\eps/4}$ was chosen so that the product of
fields $\varphi^*\varphi$ has the same dimension as the Lagrangian
density. 

Then we rewrite the Lagrangian density as a sum of three parts, 
$\mathcal{L}=\mathcal{\bar L}_0+\mathcal{\bar L}_1+\mathcal{\bar L}_2$, 
where 
\begin{align}\label{eq:L-2d}
 \mathcal{\bar L}_0 & = \Psi^\+\left(i\d_t + \frac{\sigma_3\grad^2}{2m}
 + \mu\sigma_3 + \sigma_+\phi_0 + \sigma_-\phi_0\right)\Psi \,, \\ 
 \mathcal{\bar L}_1 & = - \varphi^*\varphi 
 + \bar g\Psi^\+\sigma_+\Psi\varphi 
 + \bar g\Psi^\+\sigma_-\Psi\varphi^*, \phantom{\int}\hspace{-4.5mm} \\
 \mathcal{\bar L}_2 & = \varphi^*\varphi\,.
\end{align}
The part $\mathcal{\bar L}_0$ describes the gapped fermion
quasiparticle, whose propagator is given by
\begin{equation}
 \begin{split}
  \bar G(p_0,\p) &= \frac1{p_0^{\,2}-\bar E_\p^{\,2}+i\delta} \\
  & \qquad\quad \times
  \begin{pmatrix}
   p_0 + \ep -\mu & -\phi_0 \\
   -\phi_0 & p_0 -\ep + \mu
  \end{pmatrix},
 \end{split}
\end{equation}
with $\bar E_\p=\sqrt{(\ep-\mu)^2+\phi_0^{\,2}}$ being the usual gapped
quasiparticle spectrum in the BCS theory.

\begin{figure}[tp]
 \includegraphics[width=0.45\textwidth,clip]{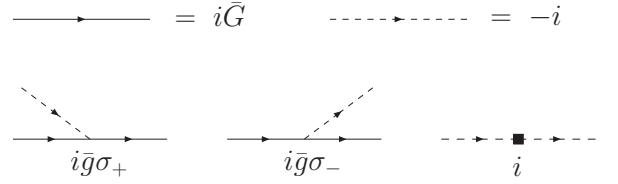}
 \caption{Feynman rules for the expansion around two spatial
 dimensions from the Lagrangian density in Eq.~(\ref{eq:L-2d}).  The
 first line gives propagators, while the second line gives vertices.
 \label{fig:feynman_rules-2d}} 
\end{figure}

The second part $\mathcal{\bar L}_1$ describes the interaction between
fermions induced by the auxiliary field $\varphi$.  The first term in
$\mathcal{\bar L}_1$ gives the propagator of the auxiliary field
$\varphi$,
\begin{equation}
 \bar D(p_0,\p) = -1, 
\end{equation} 
and the last two terms give vertices coupling two fermions with
$\varphi$.  If we did not have the part $\mathcal{\bar L}_2$, we could 
integrate out the auxiliary fields $\varphi$ and $\varphi^*$ to lead to 
\begin{equation}
 \mathcal{\bar L}_1\to \bar g^2\Psi^\+\sigma_+\Psi\,\Psi^\+\sigma_-\Psi
  =\bar g^2\psi^\dagger_\uparrow\psi^\dagger_\downarrow
  \psi_\downarrow\psi_\uparrow,
\end{equation}
which gives the contact interaction of fermions with the small coupling 
$\bar g^2\sim\bar\eps$ as depicted in Fig.~\ref{fig:scattering}.  
The vertex in the third part $\mathcal{\bar L}_2$ plays a role of a
counterterm so as to avoid double counting of a certain type of diagrams
which is already taken into $\mathcal{\bar L}_1$ as we will see below.
The Feynman rules corresponding to these Lagrangian densities are
summarized in Fig.~\ref{fig:feynman_rules-2d}. 

\subsection{Power counting rule of $\bar\eps$}
We can construct the similar power counting rule of $\bar\eps$ as in the
case of the expansion around four spatial dimensions.  Let us first
consider Feynman diagrams constructed only from $\mathcal{\bar L}_0$ and
$\mathcal{\bar L}_1$, without the vertices from $\mathcal{\bar L}_2$.
We make a prior assumption $\phi_0/\mu\sim e^{-1/\bar\eps}$, which will
be checked  later, and consider $\mu$ to be $O(1)$.  Since
$e^{-1/\bar\eps}$ is exponentially small compared to any powers of
$\bar\eps$, we can neglect the contribution of $\phi_0$ when we expand
physical observables in powers of $\bar\eps$.  Then, since each pair of
fermion and $\varphi$ vertices brings a factor of $\bar\eps$, the power
of $\bar\eps$ for a given diagram is naively $N_{\bar g}/2$, where
$N_{\bar g}$ is the number of couplings $\bar g$ from 
$\mathcal{\bar L}_1$.  However, this naive power counting does not take
into account the possibility of inverse powers of $\bar\eps$ from
loop integrals which are logarithmically divergent at $d=2$.  Let us
identify diagrams which have this divergence. 

\begin{figure}[bp]
 \includegraphics[width=0.45\textwidth,clip]{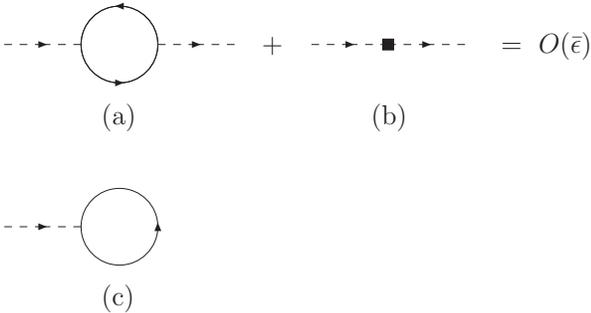}
 \caption{Two apparent exceptions of naive power counting rule of
 $\bar\eps$, [(a), (c)].  The boson self-energy diagram (a) is combined
 with the vertex from $\mathcal{\bar L}_2$ (b) to restore the naive 
 $\bar\eps$ power counting.  The condition of disappearance of the
 tadpole diagram (c) gives the gap equation to determine the value of
 condensate $\phi_0$.  \label{fig:cancel-2d}} 
\end{figure}

Consider a diagram with $L$ loop integrals, $P_\mathrm{F}$ fermion
propagators, and $P_\mathrm{B}$ internal auxiliary field lines.  Each
momentum loop integral in the ultraviolet region behaves as
\begin{equation}
 \int dp_0d\p\sim \int d\p\,\ep\sim p^{4},
\end{equation}
while each fermion propagator falls at least as $\bar G(p)\sim p^{-2}$
or faster.  Therefore, the superficial degree of divergence the diagram
has is
\begin{equation}
 \mathcal{D}=4L-2P_\mathrm{F}.
\end{equation}
Using the similar relations to Eq.~(\ref{eq:loop}), which involve the
number of external fermion (auxiliary field) lines $E_\mathrm{F(B)}$, 
\begin{equation}
 \begin{split}
  L&=P_\mathrm{F}+P_\mathrm{B}-N_{\bar g}+1,\\
  N_g&=P_\mathrm{F}+\frac{E_\mathrm{F}}2=2P_\mathrm{B}+E_\mathrm{B},
 \end{split}
\end{equation}
the superficial degree of divergence can be written as
\begin{equation}
 \mathcal{D}=4-E_\mathrm{F}-2E_\mathrm{B}.
\end{equation}
This shows that the inverse powers of $\bar\eps$ appear only in diagrams
where $E_\mathrm{F}+2E_\mathrm{B}\leq4$ is satisfied.  Moreover, from
the analytic properties of $G(p)$ in the ultraviolet region discussed in
Eq.~(\ref{eq:analytic}), one can show that there are only two skeleton
diagrams which have the $1/\bar\eps$ singularity near $d=2$.  They are
one-loop diagrams of the boson self-energy
[Fig.~\ref{fig:cancel-2d}(a)], and the $\varphi$ tadpole diagram
[Fig.~\ref{fig:cancel-2d}(c)].  We shall examine these apparent two
exceptions of the naive power counting rule of $\bar\eps$. 

The boson self-energy diagram in Fig.~\ref{fig:cancel-2d}(a) is
evaluated as 
\begin{align}
 -i\bar\Pi_\mathrm{a}(p) &= -\bar g^2\int\!\frac{dk}{(2\pi)^{d+1}}\,
 G_{11}\!\left(k+\frac p2\right)G_{22}\!\left(k-\frac p2\right) \notag\\ 
 &=i\bar g^2\int_\k \frac{\theta(\varepsilon_{\k+\frac\p2}-\mu)-\theta
 (\mu-\varepsilon_{\k-\frac\p2})}{2\ek-(p_0-\frac12\ep+2\mu+i\delta)},
\end{align}
where we have neglected the contribution of $\phi_0$. 
The integral over $\k$ is logarithmically divergent at $d=2$ and has a
pole at $\bar\eps=0$.  Thus $\bar\Pi_\mathrm{a}(p)$ is $O(1)$ instead of
$O(\bar\eps)$ according to the naive power counting.  The residue at the
pole can be computed as
\begin{align}
 \bar\Pi_\mathrm{a}(p) & = -\bar g^2\int_\k
 \frac1{2\ek-(p_0-\frac12\ep+2\mu+i\delta)} + \cdots \notag\\
 & = 1 + O(\bar\eps),
\end{align}
and is canceled out exactly by adding the vertex
$\bar\Pi_0=-1$ in $\mathcal{\bar L}_2$.  Therefore the diagram
of the type in Fig.~\ref{fig:cancel-2d}(a) should be combined with the 
vertex from $\mathcal{\bar L}_2$ in Fig.~\ref{fig:cancel-2d}(b) to
restore the naive $\bar\eps$ power counting result, i.e.,
$O(\bar\eps)$.

Similarly, the tadpole diagram in Fig.~\ref{fig:cancel-2d}(c) also
contains a $1/\bar\eps$ singularity.  The requirement that this
tadpole diagram should vanish by itself gives the gap equation to the
leading order that determines the condensate $\phi_0$ as we will see in
the succeeding subsection.


\subsection{Gap equation}
The condensate $\phi_0$ as a function of the chemical potential $\mu$ is
determined by the gap equation which is obtained by the condition of the
disappearance of all tadpole diagrams.  The leading contribution to the
gap equation is the one-loop diagram drawn in Fig.~\ref{fig:cancel-2d}(c),
which is given by 
\begin{equation}
 \bar\Xi_1 = \bar g\int\!\frac{dk}{(2\pi)^{d+1}}\,G_{21}(k)
  = i\bar g\int_\k \frac{\phi_0}{2\bar E_\k}.
\end{equation}
By changing the integration variable to $z=\ek/\mu$, we obtain 
\begin{equation}
 \begin{split}
  \bar\Xi_1 &= \frac{i\bar g\phi_0}{2\mu}
  \frac{\left(\frac{m\mu}{2\pi}\right)^{\frac d2}}
  {\Gamma\!\left(\frac d2\right)} 
  \int_0^\infty\!dz\,\frac{z^{\bar\eps/2}}{\sqrt{(z-1)^2+(\phi_0/\mu)^2}}.
 \end{split}
\end{equation}
The integration over $z$ in the dimensional regularization can be
performed to lead to 
\begin{equation}
 \bar\Xi_1 = \frac{i\bar g\phi_0}{\mu}
  \frac{\left(\frac{m\mu}{2\pi}\right)^{\frac d2}}
  {\Gamma\!\left(\frac d2\right)}\left[\ln\frac{2\mu}{\phi_0}
	    -\frac1{\bar\eps}+O(\bar\eps^2)\right].
\end{equation}
The first term $\ln2\mu/\phi_0$ originates from the singularity
around the Fermi surface as is well known as the Cooper instability, 
while the second term $1/\bar\eps$ is from the logarithmic singularity
of the $\k$ integration at $d=2$ in $\bar\Xi_1$.  Solving the gap
equation $\bar\Xi_1=0$,  we obtain the condensate as
$\phi_0=2\mu\,e^{-1/\bar\eps}$.  Note that this result is equivalent to
that obtained by the mean field BCS theory. 

\begin{figure}[tp]
 \includegraphics[width=0.25\textwidth,clip]{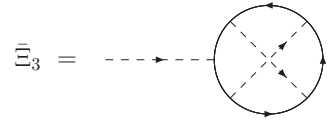}
 \caption{The tadpole diagram which gives the medium-effect
 correction to the gap equation.  \label{fig:tadpole-2d}} 
\end{figure}

It is known that the preexponential factor in the mean field result
$\phi_0=2\mu\,e^{-1/\bar\eps}$ is modified due to the effects of
medium~\cite{gorkov,heiselberg}.  In the language of the tadpole
diagrams, the corresponding modification to the gap equation comes from
the three-loop diagram $\bar\Xi_3$ depicted in
Fig.~\ref{fig:tadpole-2d}.  The diagram, which seems proportional to
$g^4\sim\bar\eps^2$, turns out to give the $O(1)$ correction to the gap
equation. 

Using the Feynman rules, the tadpole diagram in
Fig.~\ref{fig:tadpole-2d} is given by 
\begin{align}
  \bar\Xi_3
  &=-\bar g^5\int\!\frac{dk\,dp\,dq}{(2\pi)^{3d+3}}
  \Tr\!\left[\bar G\sigma_+\bar G\sigma_-
  \bar G\sigma_+\bar G\sigma_+\bar G\sigma_-\right] \notag\\
 \begin{split}
  &=-\bar g^5\int\!\frac{dk\,dp\,dq}{(2\pi)^{3d+3}}\,
  \bar G_{11}(p)\,\bar G_{22}(p) \\ &\qquad\qquad\times
  \bar G_{11}(p-k)\,\bar G_{21}(q)\,\bar G_{22}(q+k).
 \end{split}
\end{align}
Since the second term in the product,
\begin{equation}
 \bar G_{11}(p)\,\bar G_{22}(p)
  =\frac1{p_0^{\,2}-\bar E_\p^{\,2}}
  +\frac{\phi_0^{\,2}}{(p_0^{\,2}-\bar E_\p^{\,2})^2},
\end{equation}
gives only the $O(\bar\eps)$ correction to the gap equation, we can
neglect it for the current purpose.  Then we perform the integrations
over $p_0$ and $q_0$ to result in
\begin{align}
  \bar\Xi_3
  &=-\bar g^5\int\!\frac{dk\,d\p\,d\q}{(2\pi)^{3d+1}}\,
  \frac{\phi_0}{4\bar E_\p \bar E_\q}\\
  &\times\left[\frac{\bar E_\p-k_0+\varepsilon_{\p-\k}-\mu}
  {(\bar E_\p-k_0)^2-\bar E_{\p-\k}^{\,2}}
  +\frac{\bar E_\p+\varepsilon_\p-\mu}
  {(k_0+\bar E_\p)^2-\bar E_{\p+\k}^{\,2}}\right]\notag\\
  &\times \left[\frac{k_0+\bar E_\q-\varepsilon_{\q+\k}+\mu}
  {(k_0+\bar E_\q)^2-\bar E_{\q+\k}^{\,2}}
  +\frac{\bar E_\q-\varepsilon_\q+\mu}
  {(k_0-\bar E_\q)^2-\bar E_{\q-\k}^{\,2}}\right].\notag
\end{align}
Because of the factor $1/\bar E_\p \bar E_\q$ in the integrand, the
integrations over $\p$ and $\q$ are dominated around the Fermi surface
where $\varepsilon_{\p (\q)}\sim\mu$ and hence 
$\bar E_{\p (\q)}\sim\phi_0\ll\mu$.  Keeping only the dominant part in
the integrand, we can write the integral as
\begin{align}
  \bar\Xi_3
  &\simeq\bar g^5\int\!\frac{dk\,d\p\,d\q}{(2\pi)^{3d+1}}\,
  \frac{\phi_0}{4\bar E_\p \bar E_\q}\\ &\qquad \left. \times
  \frac{1}{(k_0+\varepsilon_{\p-\k}-\mu)(k_0+\varepsilon_{\q+\k}-\mu)}
  \right|_{\varepsilon_{\p(\q)}=\mu}. \notag
\end{align}
Now the integration over $k_0$ can be performed easily to lead to
\begin{equation}
 \begin{split}
  \bar\Xi_3 &= -i\frac{\bar g^5}2\phi_0
  \int_{\p\q} \frac{1}{\bar E_\p \bar E_\q}\\
  &\qquad \times \int_\k \left.
  \frac{\theta(\varepsilon_{\q+\k}-\mu)\,\theta(\mu-\varepsilon_{\p-\k})}
  {\varepsilon_{\q+\k}-\varepsilon_{\p-\k}}\right|_{\varepsilon_{\p(\q)}=\mu}.
 \end{split}
\end{equation}
If we evaluated the $\k$ integration in the second line at $d=3$, we 
would obtain the static Lindhard function representing the
medium-induced interaction. 

Here we shall evaluate $\bar\Xi_3$ at $d=2$.  Changing the
integration variables to $\ep$, $\eq$, $\ek$,
$\cos\chi_p=\hat\k\cdot\hat\p$, $\cos\chi_q=\hat\k\cdot\hat\q$, and
performing the integrations over $\ep$ and $\eq$, we obtain
\begin{align}
 \bar\Xi_3 &\simeq -i\frac{\bar g^5}2\phi_0
 \left(\frac m{2\pi}\right)^3\left(2\ln\frac\mu{\phi_0}\right)^2 \\
 &\times\int\!\frac{d\ek\,d\chi_p\,d\chi_q}{\pi^2} \left.
 \frac{\theta(\varepsilon_{\q+\k}-\mu)\,\theta(\mu-\varepsilon_{\p-\k})}
 {\varepsilon_{\q+\k}-\varepsilon_{\p-\k}}\right|_{\varepsilon_{\p(\q)}=\mu}.
 \notag
\end{align}
The range of the integrations over $\chi_p$ and $\chi_q$ are from $0$ to
$\pi$.  Since $\phi_0/\mu\propto e^{-1/\bar\eps}$, we have
$(2\ln\mu/\phi_0)^2\sim(2/\bar\eps)^2$ which cancels $\bar\eps^2$ coming
from the four vertex couplings $\bar g^4$.  Finally the remaining
integrations can be performed as follows:
\begin{widetext}
 \begin{equation}
 \begin{split}
  &\int\!\frac{d\ek\,d\chi_p\,d\chi_q}{\pi^2} \left.
  \frac{\theta(\varepsilon_{\q+\k}-\mu)\,\theta(\mu-\varepsilon_{\p-\k})}
  {\varepsilon_{\q+\k}-\varepsilon_{\p-\k}}\right|_{\varepsilon_{\p(\q)}=\mu}\\
  &=\left[\int_0^{4\mu\cos^2\!\chi_p}\!d\ek
  \int_0^{\pi/2}\!\frac{d\chi_p}{\pi}\int_0^{\pi/2}\!\frac{d\chi_q}{\pi}
  +\int_{4\mu\cos^2\!\chi_q}^{4\mu\cos^2\!\chi_p}\!d\ek\int_0^{\pi/2}\!
  \frac{d\chi_p}{\pi}\int_{\pi/2}^{\pi-\chi_p}\!\frac{d\chi_q}{\pi}\right]
  \frac{1}{2\sqrt{\mu\ek}\left(\cos\chi_p+\cos\chi_q\right)}\\
  &=\frac12\,,
 \end{split}
\end{equation}
\end{widetext}
which gives $\bar\Xi_3$ as
\begin{equation}
 \bar\Xi_3\simeq-i\bar g\phi_0\frac{m}{2\pi}.
\end{equation}

Consequently, the gap equation $\bar\Xi_1+\bar\Xi_3=0$, which receives the
$O(1)$ correction due to $\bar\Xi_3$, is modified as
\begin{equation}
 \ln\frac{2\mu}{\phi_0}-\frac1{\bar\eps}-1+O(\bar\eps)=0.
\end{equation}
The solution of the gap equation becomes
\begin{equation}
 \phi_0=2\mu\,\exp\!\left[-\frac1{\bar\eps}-1+O(\bar\eps)\right]
  =\frac{2\mu}{e}\left[1+O(\bar\eps)\right]e^{-1/\bar\eps},
\end{equation}
where the value of condensate is reduced by the factor 
$e\approx2.71828$. 
The reduction of the preexponential factor due to the medium effects
is known as the Gor'kov--Melik-Barkhudarov correction at $d=3$
theories~\cite{gorkov,heiselberg}.

\subsection{Thermodynamic quantities}
\begin{figure}[tp]
 \includegraphics[width=0.30\textwidth,clip]{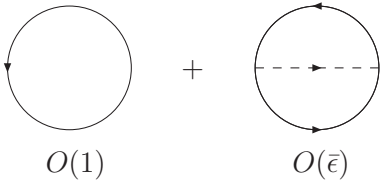}
 \caption{Vacuum diagrams contributing to the pressure up to the
 next-to-leading order in $\bar\eps$.  \label{fig:potential-2d}}  
\end{figure}

The value of the effective potential $V_\mathrm{eff}$ at its minimum
determines the pressure $P=-\bar V_\mathrm{eff}(\phi_0)$ at a given
chemical potential $\mu$.  Since the energy gain due to the
superfluidity $\phi_0^{\,2}\sim e^{-2/\bar\eps}$ is exponentially small
compared to any power series of $\bar\eps$, we can simply neglect the
contribution of $\phi_0$ to the pressure.  Up to the next-to-leading
order, the effective potential receives contributions from two vacuum
diagrams drawn in Fig.~\ref{fig:potential-2d}: fermion loops without and
with an exchange of the auxiliary field.  The one-loop diagram is $O(1)$
and given by
\begin{equation}
 \bar V_1(0) = -2\int_\p \left(\mu-\ep\right)_> 
  =-\frac{2\mu\left(\frac{m\mu}{2\pi}\right)^{\frac d2}}
  {\Gamma\!\left(\frac d2+2\right)}\equiv -P_\mathrm{free},
\end{equation}
which represents the contribution of free fermions to the pressure. 
The two-loop diagram is $O(\bar\eps)$, which represents the
density-density correlation as
\begin{equation}
 \begin{split}
  \bar V_2(0) &= -\bar g^2\int\!\frac{dp\,dq}{(2\pi)^{2d+2}}\,
  \bar G_{11}(p)\bar G_{22}(q)\\  &= -\bar g^2
  \left[\int_\p \theta(\mu-\ep)\right]^2
  =-\eps\frac{\mu\left(\frac{m\mu}{2\pi}\right)^{\frac d2}}
  {\Gamma\!\left(\frac d2+1\right)^2}. 
 \end{split}
\end{equation}
Thus we obtain the pressure up to the next-to-leading order in
$\bar\eps$ as
\begin{equation}
 P=\left(1+\bar\eps\right)P_\mathrm{free}.
\end{equation}

Accordingly, the fermion number density is given by
$N=\d P/\d\mu=\left(1+\bar\eps\right)N_\mathrm{free}$.  The Fermi
energy is obtained from the thermodynamics of free gas in $d$ spatial
dimension as 
\begin{equation}\label{eq:eF-2d}
  \eF=\frac{2\pi}m
  \left[\frac12\,\Gamma\!\left(\frac d2+1\right)N\right]^{2/d}
  =\left(1+\bar\eps\right)\mu,
\end{equation}
which yields the universal parameter of the unitary Fermi gas from the
$\bar\eps$ expansion as
\begin{equation}\label{eq:xi-2d}
 \xi=\frac\mu\eF=1-\bar\eps+O(\bar\eps^2).
\end{equation}
The $O(\bar\eps^2)$ correction to $\xi$ was recently computed to obtain
$1.52\,\bar\eps^2$~\cite{Nishida-thesis}. 

\subsection{Quasiparticle spectrum}
\begin{figure}[tp]
 \includegraphics[width=0.5\textwidth,clip]{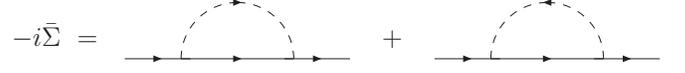}
 \caption{One-loop diagrams contributing to the fermion self-energy to 
 order $O(\bar\eps)$. \label{fig:self_energy-2d}}
\end{figure}

To the leading order in $\bar\eps$, the dispersion relation of the
fermion quasiparticle is given by 
$\omega_\mathrm{F}(\p)=\bar E_\p=\sqrt{(\ep-\mu)^{2}+\phi_0^{\,2}}$,
which has the same form as that in the mean field BCS theory. 
There exist the next-to-leading-order corrections to the fermion
quasiparticle spectrum from the one-loop self-energy diagrams
$-i\bar\Sigma(p)$ depicted in Fig.~\ref{fig:self_energy-2d}.
These corrections are only to the diagonal elements of the self-energy
and each element is evaluated as
\begin{equation}
 \begin{split}
  \bar\Sigma_{11}(p) 
  &= -i\bar g^2\int\!\frac{dk}{(2\pi)^{d+1}}\,\bar G_{22}(k)\\
  &=-\bar g^2\int_\k \theta(\mu-\ek)=-\bar\eps\mu,
 \end{split}
\end{equation}
and
\begin{equation}
 \begin{split}
  \bar\Sigma_{22}(p) 
  &= -i\bar g^2\int\!\frac{dk}{(2\pi)^{d+1}}\,\bar G_{11}(k)\\
  &=\bar g^2\int_\k \theta(\mu-\ek)=\bar\eps\mu.
 \end{split}
\end{equation}
To this order, the self-energy is momentum independent, which
effectively shifts the chemical potential due to the interaction with
the other component of fermions. 

By solving the equation
$\det[\bar G^{-1}(\omega,\p)-\bar\Sigma]=0$ in terms of $\omega$, 
the dispersion relation of the fermion quasiparticle up to the
next-to-leading order is given by 
\begin{equation}
  \omega_\mathrm{F}(\p)=\sqrt{(\ep-\mu-\bar\eps\mu)^2+\phi_0^{\,2}}. 
\end{equation}
The minimum of the dispersion curve is located at a nonzero value of
momentum, $|\p|=(2m\varepsilon_0)^{1/2}$, where
\begin{equation}\label{eq:e_0-2d}
 \varepsilon_0=\left(1+\bar\eps\right)\mu.
\end{equation}
The location of the minimum coincides with the Fermi energy in
Eq.~(\ref{eq:eF-2d}), $\varepsilon_0=\eF$, in agreement with the
Luttinger theorem~\cite{luttinger}.  The energy gap $\Delta$ of the
fermion quasiparticle is given by the condensate, 
\begin{equation}\label{eq:gap-2d}
 \Delta=\phi_0 =\frac{2\mu}{e}\,e^{-1/\bar\eps}.
\end{equation}

\subsection{Extrapolation to $\bar\eps$=1}
Now we discuss the extrapolation of the series expansion over
$\bar\eps=d-2$ to the physical case in three spatial dimensions.  In
contradiction to the case of the $\eps=4-d$ expansion, the coefficients
of the $O(\bar\eps)$ corrections are not small compared to the leading
terms.  If we naively extrapolate the leading and next-to-leading-order
results for $\xi$ in Eq.~(\ref{eq:xi-2d}), $\Delta$ in
Eq.~(\ref{eq:gap-2d}), and $\varepsilon_0$ in Eq.~(\ref{eq:e_0-2d}) to
$\bar\eps=1$, we would have
\begin{equation}
 \xi \approx 0, \qquad \frac{\Delta}{\mu}\approx 0.271, 
  \qquad \frac{\varepsilon_0}{\mu}\approx 2, 
\end{equation}
which are not as good as the extrapolations in the expansions over
$\eps$ in Eq.~(\ref{eq:extrapolation}).  This may be related to the fact
that the series expansion over $\bar\eps=d-2$ is an asymptotic series
and is not Borel summable.  Thus, instead of naively extrapolating the
$\bar\eps$ expansions to $d=3$, we use them as boundary conditions to
improve the series summation of the $\eps$ expansions in
Sec.~\ref{sec:matching}.

\subsection{Large-order behavior}
Here we show that the series expansion over $\bar\eps=d-2$ is not
convergent, but an asymptotic series.  Its physical reason is obvious:
The unitary Fermi gas at $d\leq2$ is just a free gas while the ground
state at $d>2$ is the superfluid.  Therefore, the radius of convergence
in the expansion around two spatial dimensions is zero.  In the language 
of the perturbation theory developed in this section, the asymptotic
nature of the expansion over $\bar\eps=d-2$ can be understood by the
existence of a type of diagrams which grows as $n!$ by itself at order
$\bar\eps^n$.  Such a factorial contribution originates from the low
momentum integration region which resembles the \textit{infrared
renormalon} in the relativistic field theories~\cite{Parisi,David}.

\begin{figure}[tp]
 \includegraphics[width=0.45\textwidth,clip]{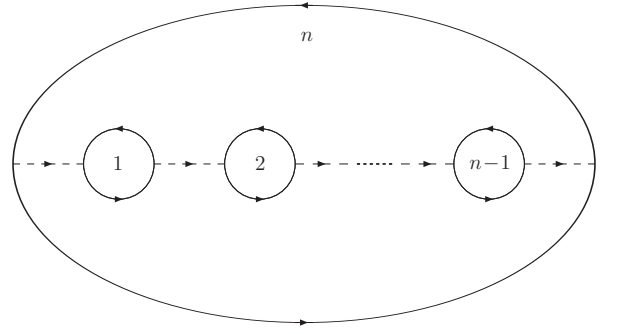}
 \caption{An $n$th order diagram at $d=2$ which contributes to the
 pressure as $n!$ by itself.  The countervertex $-i\bar\Pi_0=i$ for
 each bubble diagram is understood implicitly. \label{fig:renormalon}} 
\end{figure}

The $n+1$-loop diagram contributing to the effective potential as $n!$
at $O(\bar\eps^n)$ is depicted in Fig.~\ref{fig:renormalon}, which is
written as
\begin{align}
 \bar V_{n+1} &= \frac in\int\!\frac{dp}{(2\pi)^{d+1}} \label{eq:V_n-bar}\\
 &\times\left[1+\bar g^2\int_\k
 \frac{\theta(\varepsilon_{\k+\frac\p2}-\mu)-\theta
 (\mu-\varepsilon_{\k-\frac\p2})}{2\ek-(p_0-\frac12\ep+2\mu+i\delta)}
 \right]^n, \notag
\end{align}
where $k_0$ integrations in each bubble diagram are already performed. 
Note that $+1$ in the bracket comes from the countervertex
$-\bar\Pi_0=1$.  Since the $n!$ contribution comes from the infrared
region of $p$, we concentrate on the integration region where
$|p_0|\ll\mu$ and $\ep\ll\mu$.  Rotating the contour of the $p_0$
integration to the imaginary axis, $p_0\to i\tilde p_0$ (Wick rotation),
the $\k$ integration in the bracket can be evaluated as
\begin{equation}
 \begin{split}
  &1+\bar g^2\int_\k \frac{\theta(\varepsilon_{\k+\frac\p2}-\mu)-\theta
  (\mu-\varepsilon_{\k-\frac\p2})}{2\ek-(p_0-\frac12\ep+2\mu+i\delta)} \\
  &= -\bar\eps\,\ln\!\left(\frac{|\tilde p_0|}{4\mu}
  +\sqrt{\left(\frac{\tilde p_0}{4\mu}\right)^2+\frac{\ep}{4\mu}}\right)
  +\cdots.
 \end{split}
\end{equation}
Here $\cdots$ includes higher order terms in $\bar\eps$ and regular
dependences on $\tilde p_0$ and $\ep$.  The logarithmic dependence on 
$\tilde p_0$ and $\ep$ appears due to the singularity of the $\k$
integration near the Fermi surface $\ek\sim\mu$ in the limit $p\to0$.
Such a logarithmic term dominates the $p$ integration when $p$ is
small.  Collecting the logarithmic term from each bracket, the $\tilde
p_0$ and $\p$ integrals in Eq.~(\ref{eq:V_n-bar}) at their low momentum
regions contribute to $\bar V_{n+1}$ as
\begin{align}
 \bar V_{n+1} &\sim\frac1n\int_0d\tilde p_0d\p
  \left[-\bar\eps\,\ln\!\left(\frac{|\tilde p_0|}{4\mu}
  +\sqrt{\left(\frac{\tilde p_0}{4\mu}\right)^2
  +\frac{\ep}{4\mu}}\right)\right]^n \notag\\
  &\sim\left(\frac{\bar\eps}3\right)^n\Gamma(n),
\end{align}
which behaves as the factorial of $n$.  

Because the large-order contribution grows up with the same sign, it
produces the singularity on the real positive axis of the Borel
transform.  Thus the expansion over $\bar\eps=d-2$ is not Borel summable
and its Borel transformation will have the inevitable ambiguity $\sim
e^{-3/\bar\eps}$, indicating that the ground state where the expansion
is performed is unstable.  Finally, it is important to note that the
discussion here is not applicable to the expansion around $d=4$ because
the singularity near the Fermi surface is cured by the existence of the
large condensate $\phi_0\gg\mu$.  In the Appendix~\ref{sec:renormalon},
we explicitly demonstrate that the diagrams in Fig.~\ref{fig:renormalon}
do not yield a factorial growth in the coefficients of the $\eps$
expansion.

\section{Matching of expansions around four and two spatial dimensions 
 \label{sec:matching}}
As we have mentioned previously, we shall match the two expansions
around $d=4$ and $d=2$, which were studied in Secs.~\ref{sec:4d} and
\ref{sec:2d}, respectively, in order to extract results at $d=3$.  
We use the results around two spatial dimensions as boundary
conditions which should be satisfied by the series summations of the
expansions over $\eps=4-d$.  Because we do not yet have a precise
knowledge of the large-order behavior of the expansion around four
spatial dimensions, we assume its Borel summability and employ Pad\'e
approximants. 

Let us demonstrate the matching of the two expansions by taking $\xi$ as
an example.  In Ref.~\cite{nussinov04}, the linear interpolation between
the exact values at $d=2$ $(\xi=1)$ and $d=4$ $(\xi=0)$ was discussed to
yield $\xi=0.5$ at $d=3$.  Now we have series expansions around these
two exact limits.  The expansion of $\xi$ in terms of $\eps=4-d$ is
obtained in Eq.~(\ref{eq:xi}).  Assuming the Borel summability of the
$\eps$ expansion, we write $\xi$ as a function of $\eps$ in the form of
the Borel transformation, 
\begin{equation}\label{eq:borel}
 \xi(\eps)=\frac{\eps^{3/2}}2
 \exp\!\left(\frac{\eps\ln\eps}{8-2\eps}\right)
 \int_0^\infty\!dt\, e^{-t}B_\xi(\eps t),
\end{equation}
where we factorized out the trivial nonanalytic dependences on $\eps$
explicitly.  $B_\xi(t)$ is the Borel transform of the power series in
$\xi(\eps)$, whose Taylor coefficients at origin are given from the
$\eps$ expansion of $\xi$ as 
\begin{equation}\label{eq:borel_sum}
 B_\xi(t)=1 - \left(3C -\frac54 (1-\ln2)\right) t +\cdots.
\end{equation}

In order to perform the integration over $t$ in Eq.~(\ref{eq:borel}),
the analytic continuation of the Borel transform $B_\xi(t)$ to the real 
positive axis of $t$ is necessary.  Here we employ the Pad\'e
approximant, where $B_\xi(t)$ is replaced by the following rational
functions:
\begin{equation}
 B_\xi(t) = \frac{1+p_1 t+\cdots+p_M t^M}{1+q_1 t+\cdots+q_N t^N}\,.
\end{equation}
From Eq.~(\ref{eq:borel_sum}), we require that the Pad\'e approximants
satisfy $p_1-q_1=-3C+\frac54 (1-\ln2)$.  Furthermore, we incorporate the
results around two spatial dimensions in Eq.~(\ref{eq:xi-2d}) by
imposing 
\begin{equation}
 \xi(2-\bar\eps)=1-\bar\eps+\cdots
\end{equation}
on the Pad\'e approximants as a boundary condition.  Since we have three
known coefficients from the two expansions, the Pad\'e approximants
$[M/N]$ satisfying $M+N=3$ are possible.  Since we could not find a
solution satisfying the boundary condition $\xi(2-\bar\eps)=1-\bar\eps$
for $[M/N]=[2/1]$, we adopt three other Pad\'e approximants with
$[M/N]=[3/0],\,[1/2],\,[0/3]$, whose coefficients $p_m$ and $q_n$ are
determined uniquely by the above conditions. 

\begin{figure}[tp]
 \includegraphics[width=0.45\textwidth,clip]{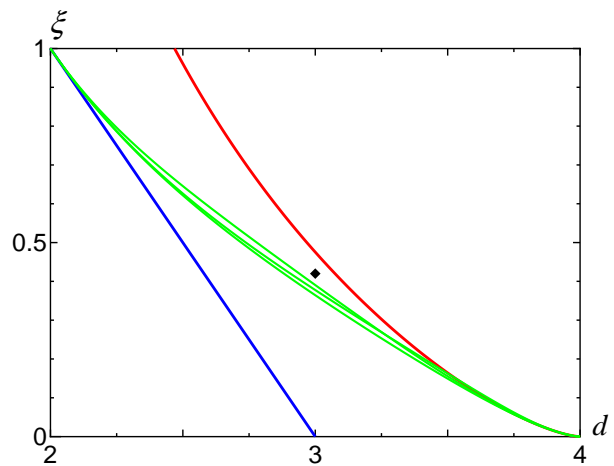}
 \caption{(Color online) The universal parameter $\xi$ as a function of
 the spatial dimensions $d$.  The upper solid curve is from the
 expansion around $d=4$ in Eq.~(\ref{eq:xi}), while the lower solid line
 is from the expansion around $d=2$ in Eq.~(\ref{eq:xi-2d}).  The middle
 three curves show the different Borel-Pad\'e approximations connecting
 the two expansions.  The diamond indicates the result $\xi\approx0.42$
 from the Monte Carlo simulation~\cite{Carlson:2005kg}. \label{fig:xi}}
\end{figure}

Figure \ref{fig:xi} shows the universal parameter $\xi$ as a function of
the spatial dimension $d$.  The middle three curves show $\xi$ in the
different Pad\'e approximants connecting the two expansions around $d=4$
and $d=2$.  These Borel-Pad\'e approximations give $\xi=0.391$,
$0.364$, and $0.378$ at $d=3$,  
which are small compared to the naive extrapolation of the
$\eps$ expansion to $d=3$ $(\xi\to0.475)$. 

The Pad\'e approximant above, however, almost certainly needs serious
modification, since we can show that in the expansion of $\xi(\eps)$,
there exist nonanalytic terms at sufficiently higher orders in $\eps$,
\begin{align}
 \begin{split}
  \xi(\eps) &= \frac{\eps^{3/2}}2 
  \exp\left(\frac{\eps\ln\eps}{8-2\eps}\right)\\ &\quad\times
  \left(1 - 0.0492\eps + \#\,\eps^2 + \#\,\eps^3\ln\eps + \cdots\right).
 \end{split} 
\end{align}
The nonanalytic term at $O(\eps^3)$ compared to the leading order
originates from a resummation of boson self-energies to avoid infrared
singularities. 
An understanding of the analytic structure of high-order terms in the
perturbation theory around $d=4$ is currently lacking.

\section{Summary and concluding remarks \label{sec:summary}}
We have developed the systematic expansion for the Fermi gas near the 
unitarity limit treating the dimensionality of space $d$ as close to 
four.  To the leading and next-to-leading orders in the expansion over 
$\eps=4-d$, the thermodynamic quantities and the fermion quasiparticle
spectrum were calculated as functions of the binding energy $\eb$ of the
two-body state.  Results for the physical case of three spatial
dimensions were obtained by extrapolating the series expansions to
$\eps=1$.  The pressure, energy density, and chemical potential in the
unitary regime at fixed density were found to be
\begin{align}
 \frac{P}{\eF N} &= \frac2{d+2}\,\xi -\frac\eps4\frac\eb{2\eF}, \\
 \frac{E}{\eF N} &= \frac{d}{d+2}\,\xi 
 - \frac\eb{2\eF}\left(1+\frac\eps4+\frac{\eps\ln\eps}4
 -\frac\eps2\ln\frac\eb\eF\right), 
 \phantom{\frac{\frac\int\int}{\frac\int\int}}\hspace{-4.5mm} \\ 
 \frac{\mu}{\eF} &= \xi - \frac\eb{2\eF}\left(1+\frac\eps2
 +\frac{\eps\ln\eps}4-\frac\eps2\ln\frac\eb\eF\right),
\end{align}
where 
\begin{equation}
 \xi =\frac{\eps^{3/2}}2+\frac{\eps^{5/2}}{16}\ln\eps
  -0.0246\,\eps^{5/2}+O(\eps^{7/2}) \approx0.475
\end{equation}
is the universal parameter of the unitary Fermi
gas.  The fermion quasiparticle spectrum around its minimum in the
unitarity limit was given by the form
$\omega_\mathrm{F}(\p)\simeq\sqrt{(\ep-\varepsilon_0)^2+\Delta^2}$ with
the energy gap 
\begin{equation}
 \frac\Delta\mu=\frac2\eps-0.691+O(\eps)\approx1.31
\end{equation}
and the location of the minimum of the dispersion curve 
\begin{equation}
 \frac{\varepsilon_0}\mu=2+O(\eps)\approx2.
\end{equation}  
Although we have only the first two terms in the $\eps$ expansion, these
extrapolated values give reasonable results consistent with the Monte
Carlo simulations or the experimental measurements.  Furthermore, the
next-to-leading-order corrections are not too large even at $\eps=1$,
which suggests that the picture of the unitary Fermi gas as a collection
of weakly interacting fermionic and bosonic quasiparticles may be a
useful starting point even in three spatial dimensions. 

We have also formulated the systematic expansion for the unitary Fermi
gas around two spatial dimensions.  We used the results around $d=2$ as
boundary conditions which should be satisfied by the series summations
of the expansion over $\eps=4-d$.  The simple Borel-Pad\'e
approximations connecting the two expansions yielded $\xi=0.378\pm0.013$
at $d=3$, which is small compared to the naive extrapolation of the
$\eps$ expansion (Fig.~\ref{fig:xi}).  In order for the accurate
determination of $\xi$ at $d=3$, the precise knowledge on the
large-order behavior of the expansion around four spatial dimensions as
well as the calculation of higher order corrections would be desirable.
Once this information becomes available, a conformal mapping technique,
if applicable, will further improve the series summations~\cite{justin}. 

The phase structure of the polarized Fermi gas in the unitary regime has
been studied based on the $\eps$ expansion.  We found the splitting 
point, where the minimum of the fermion dispersion curve sits exactly at
zero momentum, located on the BEC side of the unitarity point where
$a\kF\approx1$  and $\mu/\eF\approx-0.5$.  Then the gapless superfluid
phase and the superfluid phase with the spatially varying condensate
were shown to exist between the gapped superfluid phase and the
polarized normal phase in a certain range of the binding energy.  Our
study gives a microscopic foundation to the phase structure around the
splitting point which has been proposed using the effective field
theory~\cite{son05}.  Moreover, we found the gapless phase with the
spatially varying condensate around $d=4$ exists only at
$0.494\lesssim(a\kF)^{-1}\lesssim1$ and terminates near the unitarity
limit in contrast to the conjectured phase diagram where that phase is
connected to the FFLO phase in the BCS limit~\cite{son05}.  Our result
suggests that the superfluid phase with the spatially varying condensate
existing in the unitary regime may be separated from the FFLO phase in
the BCS regime.  Further study will be worthwhile to confirm this
possibility. 

We believe that the $\eps$ expansion is not only theoretically 
interesting but also provides us an useful analytical tool to
investigate the properties of the unitary Fermi gas.  As far as we know,
this is the only systematic expansion for the unitary Fermi gas at zero
temperature that exists at this moment.  The application of the $\eps$
expansion to study the thermodynamics of the unitary Fermi gas at finite
temperature will be reported elsewhere~\cite{finite-T}.

{\itshape Note added.} 
Renormalization group argument to support our results appeared after we
submitted our manuscript.  Reference \cite{Sachdev} studied a
renormalization group flow equation and found the fixed point describing
the unitarity limit to approach the noninteracting Gaussian fixed point
in the limit $d=4$ or $d=2$.  Therefore, slightly below $d=4$ or above
$d=2$, the finite density system associated with the fixed point is
weakly coupled.  The small effective coupling in our Eq.~(\ref{eq:g}) or
Eq.~(\ref{eq:g-bar}) was correctly reproduced by the renormalization 
group fixed point~\cite{Sachdev}.

\begin{acknowledgments}
 Y.\,N.\ was supported by the Japan Society for the Promotion of Science 
 for Young Scientists.  This work was supported, in part, by DOE Grant
 No.\ DE-FG02-00ER41132.
\end{acknowledgments}

\onecolumngrid\appendix*
\section{Cancellation of renormalons at $d=4-\eps$\label{sec:renormalon}}
Here we discuss the \textit{ultraviolet
renormalon}~\cite{Gross-Neveu,Lautrup,tHooft}, which is naively present
in the diagram depicted in Fig.~\ref{fig:renormalon} and produces a $n!$
contribution at order $\eps^n$ from the large momentum integration
region of a single diagram.  The purpose of this Appendix is to show
that the $n!$ contribution cancels with subleading contributions from
lower-order diagrams in the expansion over $\eps=4-d$.  The $n+1$-loop
diagram in Fig.~\ref{fig:renormalon} can be written as
\begin{equation}
 V_{n+1}=\frac in\int\!\frac{dp}{(2\pi)^{d+1}}
  \left[D(p)\left\{\Pi_0(p)+\Pi_\mathrm{a}(p)\right\}\right]^n,
\end{equation}
where $D(p)$, $\Pi_0(p)$, and $\Pi_\mathrm{a}(p)$ are, respectively,
defined in Eqs.~(\ref{eq:D}), (\ref{eq:Pi_0}), and (\ref{eq:Pi_a}). 
Introducing these definitions and integrating over $p_0$, we obtain the
following expression for $V_{n+1}$:
\begin{equation}\label{eq:V_n}
 \begin{split}
  V_{n+1} &= -\frac{g^2}4\int_{\p\k}
  \frac{(E_{\k-\frac\p2}-\varepsilon_{\k-\frac\p2})
  (E_{\k+\frac\p2}-\varepsilon_{\k+\frac\p2})}
  {E_{\k-\frac\p2}E_{\k+\frac\p2}
  \left(E_{\k-\frac\p2}+E_{\k+\frac\p2}+\frac\ep2\right)}\\
  &\qquad\qquad\times
  \left[1+\frac{g^2}{E_{\k-\frac\p2}+E_{\k+\frac\p2}+\frac\ep2}
  \int_\l \frac1{4E_{\l-\frac\p2}E_{\l+\frac\p2}}
  \right.\\ &\qquad\qquad\qquad\quad\left.\times
  \left\{\frac{(E_{\l-\frac\p2}+\varepsilon_{\l-\frac\p2})
  (E_{\l+\frac\p2}+\varepsilon_{\l+\frac\p2})}
  {E_{\l-\frac\p2}+E_{\l+\frac\p2}+E_{\k-\frac\p2}+E_{\k+\frac\p2}}
  +\frac{(E_{\l-\frac\p2}-\varepsilon_{\l-\frac\p2})
  (E_{\l+\frac\p2}-\varepsilon_{\l+\frac\p2})}
  {E_{\l-\frac\p2}+E_{\l+\frac\p2}-E_{\k-\frac\p2}-E_{\k+\frac\p2}}
  \right\}\right]^{n-1}.
 \end{split}
\end{equation}
Now we consider the $\l$ integration in the bracket.  Since the $\l$
integration contains a logarithmic divergence at $d=4$, we subtract and
add its divergent piece as
\begin{equation}\label{eq:subtraction}
 \begin{split}
  &1+\frac{g^2}{E_{\k-\frac\p2}+E_{\k+\frac\p2}+\frac\ep2}
  \int_\l \frac1{4E_{\l-\frac\p2}E_{\l+\frac\p2}}
  \biggl\{\ \cdots\ \biggr\}\\
  &\qquad =\frac{g^2}{E_{\k-\frac\p2}+E_{\k+\frac\p2}+\frac\ep2}
  \int_\l \left[
  \frac1{4E_{\l-\frac\p2}E_{\l+\frac\p2}}\biggl\{\ \cdots\ \biggr\} 
  -\frac{1}{2\el+E_{\k-\frac\p2}+E_{\k+\frac\p2}+\frac\ep2}\right]\\
  &\qquad\quad+1+\frac{g^2}{E_{\k-\frac\p2}+E_{\k+\frac\p2}+\frac\ep2}
  \int_\l \frac{1}{2\el+E_{\k-\frac\p2}+E_{\k+\frac\p2}+\frac\ep2}.
 \end{split}
\end{equation}
The $\l$ integration in the second line becomes finite at $d=4$ and does
not produce singular logarithmic terms.  So we concentrate on the $\l$
integration in the last line, which can be evaluated in the dimensional
regularization as
\begin{equation}
 1+\frac{g^2}{E_{\k-\frac\p2}+E_{\k+\frac\p2}+\frac\ep2}
  \int_\l \frac{1}{2\el+E_{\k-\frac\p2}+E_{\k+\frac\p2}+\frac\ep2}
  = 1+\frac\eps2\,\Gamma\!\left(1-\frac d2\right)
  \left(\frac{E_{\k-\frac\p2}+E_{\k+\frac\p2}
   +\frac\ep2}{2\phi_0}\right)^{-\frac\eps2}.
\end{equation}
Then $V_{n+1}$ in Eq.~(\ref{eq:V_n}) becomes
\begin{equation}\label{eq:V_n2}
 V_{n+1} = -\frac{g^2}4\int_{\p\q} \frac{(\Ep-\ep)(\Eq-\eq)}
  {\Ep\Eq\left(\Ep+\Eq+\frac{\varepsilon_{\p-\q}}2\right)}
  \left[1+\frac\eps2\,\Gamma\!\left(-1+\frac\eps2\right)
   \left(\frac{\Ep+\Eq+\frac{\varepsilon_{\p-\q}}2}{2\phi_0}\right)^{-\frac\eps2}
   +\mathrm{\,regular\ terms\,}\right]^{n-1},
\end{equation}
where the integration valuables are redefined to be $\k-\frac\p2\to\p$
and $\k+\frac\p2\to\q$.  ``Regular terms'' coming from the second line
in Eq.~(\ref{eq:subtraction}) vanish in the
limit $\p\to\infty$ or $\q\to\infty$. 

If one expanded $V_{n+1}$ in powers of $\eps$ and picked up the
logarithmic terms, one would find the $\p$ integration from its large
momentum region $\ep\gg\phi_0,\eq$ gives the following contribution to
the effective potential at $O(\eps^n)$:
\begin{equation}
 \begin{split}
  V_{n+1} &= -\frac{g^2}4\int_{\p\q} \frac{(\Ep-\ep)(\Eq-\eq)}
  {\Ep\Eq\left(\Ep+\Eq+\frac{\varepsilon_{\p-\q}}2\right)}
  \left[\frac\eps2\,\ln\!
  \left(\frac{\Ep+\Eq+\frac{\varepsilon_{\p-\q}}2}{2\phi_0}\right)
  +\cdots\right]^{n-1}\\
  &\sim \eps\int\!d\ep\left(\frac1\ep\right)^2
  \left(\frac\eps2\ln\ep\right)^{n-1} 
  \sim \left(\frac\eps2\right)^n \Gamma(n),
 \end{split}
\end{equation}
which grows as a factorial of $n$.  Whether such a $n!$ contribution at
$O(\eps^n)$ really survives and dominates the large-order behavior of
the perturbative expansion is a subtle problem.  Actually in our case,
we can show that the factorial contribution of $V_{n+1}$ at $O(\eps^n)$
cancels because the lower-order diagrams $V_{i\leq n}$ also produce $n!$
contributions to $O(\eps^n)$ with different signs.

In order to see the cancellation explicitly, we sum up $V_{n+1}$ in
Eq.~(\ref{eq:V_n2}) over $n$ to result in
\begin{equation}
 \sum_{n=1}^\infty V_{n+1}
  = \frac{g^2}4\int_{\p\q} \frac{(\Ep-\ep)(\Eq-\eq)}
  {\Ep\Eq\left(\Ep+\Eq+\frac{\varepsilon_{\p-\q}}2\right)}
  \left[\frac\eps2\,\Gamma\!\left(-1+\frac\eps2\right)
   \left(\frac{\Ep+\Eq+\frac{\varepsilon_{\p-\q}}2}{2\phi_0}\right)^{-\frac\eps2}
   +\mathrm{\,regular\ terms\,}\right]^{-1}. 
\end{equation}
From the large momentum region of the $\p$ integration, we obtain  
\begin{equation}
  \sum_n V_{n+1}
  \sim \eps\int\!d\ep\left(\frac1\ep\right)^{2-\frac\eps2} 
  \sim\sum_n\left(\frac\eps2\right)^n,
\end{equation}
where the $n!$ contribution at $O(\eps^n)$ has been canceled with the
subleading contributions from the lower-order diagrams.  Since the
instanton contribution to the large orders, even if it exists, typically
has alternating signs, the possibility of the Borel summability of the
$\eps=4-d$ expansion, which was assumed in Sec.~\ref{sec:matching},
remains open.

\twocolumngrid

\end{document}